# Nanoscopic Interfacial Hydrogel Viscoelasticity Revealed from Comparison of Macroscopic and Microscopic Rheology

Robert F. Schmidt [a),t], Henrik Kiefer [b),t], Robert Dalgliesh [c)], Michael Gradzielski [a)], Roland R. Netz [*,b)]

a) Stranski-Laboratorium für Physikalische und Theoretische Chemie, Technische Universität Berlin, Strasse des 17. Juni 124, 10623 Berlin, Germany

b) Fachbereich Physik, Freie Universität Berlin, Arnimallee 14, 14195 Berlin, Germany

c) STFC, ISIS, Rutherford Appleton Laboratory, Chilton, Oxfordshire OX11 0QX, United Kingdom

t both authors contributed equally to this work and share the first authorship

**Abstract**

Deviations between macrorheological and particle-based microrheological measurements are often considered a nuisance and neglected. We study aqueous poly(ethylene oxide) (PEO) hydrogels for varying PEO concentrations and chain lengths that contain microscopic tracer particles and show that these deviations in fact reveal the nanoscopic viscoelastic properties of the particle-hydrogel interface. Based on the transient Stokes equation, we first demonstrate that the deviations are not due to finite particle radius, compressibility or surface-slip effects. Small-angle neutron scattering rules out hydrogel heterogeneities. Instead, we show that a generalized Stokes-Einstein relation, accounting for a nanoscopic interfacial shell around tracers with viscoelastic properties that significantly deviate from bulk, consistently explains our macrorheological and microrheological measurements. The extracted shell diameter is comparable with the PEO end-to-end distance, indicating the importance of dangling chain ends. Our methodology reveals the nanoscopic interfacial rheology of hydrogels and is generally applicable to different kinds of viscoelastic fluids and particles.

Soft matter materials are generally viscoelastic, meaning that they exhibit viscous, elastic or intermediate response to external perturbations, depending on the response time. In macrorheology, a macroscopic amount of material is deformed by applying strain or stress, and the resulting force or displacement response is measured, respectively.[1] A common macrorheological technique is oscillatory shear rheology, where the sample is subject to an oscillating shear strain, and the resulting oscillating shear stress is measured, yielding the complex modulus $G^*$ as a function of frequency. In contrast, in microrheology, the viscoelastic behavior of the sample is extracted from the active or passive motion of dispersed microscopic tracer particles.[2–4] Microrheology offers several advantages over macrorheology, such as smaller sample volume, the ability to probe locally in spatially heterogeneous samples, and access to much higher frequencies.

Ideally, one would like to combine macro- and microrheological techniques and obtain the viscoelastic sample response over a comprehensive frequency range, for which one needs to accurately extract the viscoelastic modulus from the tracer-particle dynamics. This is accomplished by the generalized Stokes-Einstein relation (GSER), which connects the macroscopic sample viscoelasticity to the frequency-dependent friction experienced by a tracer particle.[5,6] Because of its importance for the understanding of soft-matter dynamics, the GSER has been the subject of numerous experimental and theoretical investigations.[7–15] Several studies have compared macro- and microrheological measurements on the same sample.[5,16–19] Using the GSER for the conversion of the microrheology data, the reported agreement of the complex modulus $G^*$ in the overlap frequency range is typically rather good, however, upon closer inspection, it is evident that macro- and microrheological data exhibit systematic deviations, in the sense that microrheology experiments show enhanced or reduced viscoelastic response compared to macrorheology, depending on specificities of the sample and the tracer particles.[16,17,20]

This is where our paper comes in: We show that the experimentally determined deviations between macro- and microrheological spectra for a synthetic polymeric hydrogel reveal the effect of polymer-particle interactions on the effective hydrogel viscoelasticity around the probe particles. We employ semi-dilute aqueous solutions of linear poly(ethylene oxide) (PEO) polymers, which are hydrogels with physical crosslinks due to polymer chain entanglements[21–24] and constitute ideal model systems because of their simple structure and reproducible properties.[16,18,25–28] We tune the PEO hydrogel viscoelasticity by changing both PEO concentration and chain length.

The GSER has been argued to hold for homogeneous and incompressible samples[5,6] and in the absence of slip on the tracer-particle surface.[29] In fact, finite compressibility of the viscoelastic sample, slip effects and finite tracer particle size can be exactly accounted for by the solution of the transient Stokes equation for a viscoelastic fluid in spherical geomery,[30] but does not explain the deviations between our macro- and microrheology hydrogel data, as shown below. The effect of sample inhomogeneity is more subtle: A hydrogel, i.e., a dilute entangled polymer solution, is structurally characterized by its mesh size.[31] For tracer particles significantly larger than the mesh size, the hydrogel can be considered homogeneous on the characteristic particle length scale and the particles probe the macroscopic hydrogel viscosity. Particles much smaller than the mesh size can diffuse through the hydrogel meshes and are subject to the solvent viscosity, unless they are strongly attracted to the polymers making up the hydrogel.[32,33] The intermediate situation, if the particle size is of the order of the hydrogel inhomogeneity, characterized by the mesh size, constitutes an immensely complex problem.[34,35] In our experiments, the tracer particles are significantly larger than the hydrogel mesh size, as determined from small-angle neutron scattering (SANS) measurements, so we can confidently assume that the particles probe the macroscopic hydrogel viscoelasticity. Yet, there is another effect that intrinsically differentiates macro- from microrheological data and has hitherto not been studied in detail: Any tracer-particle material will interact attractively or repulsively with the hydrogel polymer and thereby induce polymer adsorption or depletion.[36–38] As a consequence, the effective hydrogel viscoelasticity in the vicinity of the particle surface will differ from its bulk value. By using a simple shell model for the hydrogel viscoelastic properties,[39,40] we demonstrate in this paper that we can not only explain the commonly observed deviation between macro- and microrheological data but also derive the effective viscosity in the hydrogel interfacial layer from these deviations.

**Macrorheological viscoelastic spectra of PEO solutions**. Frequency sweeps on poly(ethylene oxide) (PEO) solutions, which are viscoelastic in the semi-dilute regime (see Supporting Information (SI) Section S1, for details), were performed for varying polymer concentration $c$ and chain lengths (i.e. molecular weights) with a strain amplitude of $\gamma_0 = 5$ % and angular frequencies between 0.1 and 100 rad/s (see SI Sections S2 for sample preparation and S3 for experimental details). The results in Figure 1 demonstrate that the elastic $G'$ and viscous $G''$ moduli increase with concentration and chain length. The low-frequency plateau of $G'$ for the low-viscosity samples is a measurement artefact due to phase-angle uncertainties and expected for samples with low-torque signals.[20] For 1 MDa PEO (Figure 1A), all samples are predominantly viscous since $G'' > G'$ for all concentrations and frequencies except for the highest concentrated 4% sample, where we see a crossover at very high frequencies. The inverse crossover frequency $\omega_0$ indicates a balance between entanglement and disentanglement dynamics and defines the effective relaxation time $\tau_0 = 2\pi/\omega_0$.[41] With increasing concentration $\omega_0$, indicated by arrows in Figure 1B, shifts to lower frequencies. For the 4 MDa PEO (Figure 1C), on the other hand, $G'$ dominates for most concentrations and frequencies, indicating that these samples behave predominantly elastically. Our samples thus cover the full range of viscoelastic behavior. In SI Section S4 it is shown that the frequency dependence of $G'$ and $G''$ is well described by the fractional Maxwell model, which features power-law spectral behaviour.[42]

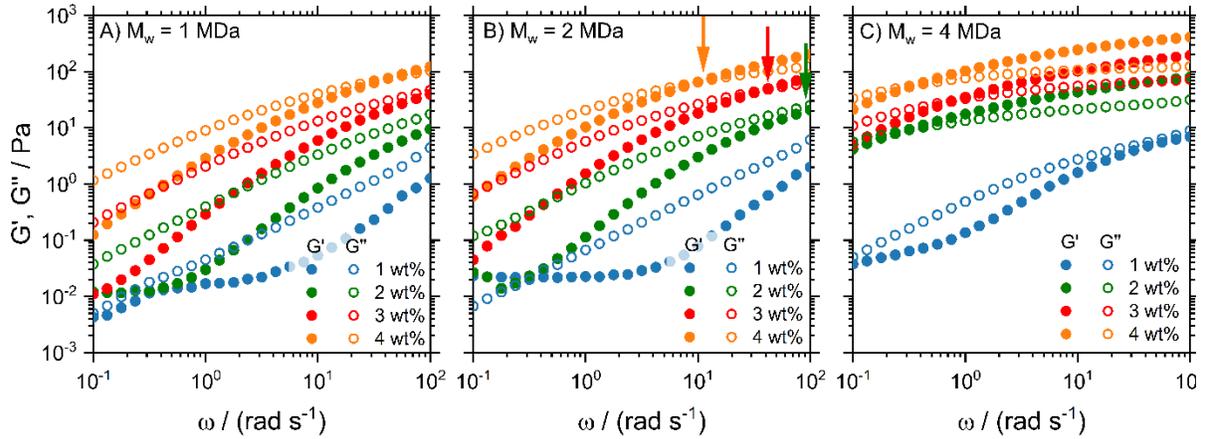

**Figure 1**. Storage ($G'$) and loss ($G''$) moduli from macrorheological oscillatory frequency sweeps for PEO solutions with different concentrations and molecular weights of (A) 1 MDa, (B) 2 MDa and (C) 4 MDa. The vertical arrows in (B) indicate the crossover frequency $\omega_0$.

**Microrheological viscoelastic spectra**. Microrheological experiments using dynamic light scattering (DLS) were performed on the same PEO samples that contain polystyrene (PS) tracer particles with hydrodynamic diameters of 68.8 (referred to as PS-69), 109.3 (PS-109), and 192.0 nm (PS-192). The DLS measurements yield the intensity auto-correlation function $g^{(2)}(\tau)$, which is converted into the mean-squared displacement (MSD) $\langle\Delta r^2(\tau)\rangle$ shown in Figure 2A-C. Only the highly viscous 4 MDa samples for 3 and 4 wt% exhibit slight deviations among different spatial measurement positions caused by the long relaxation times in these systems (for details and additional data see SI Section S5).

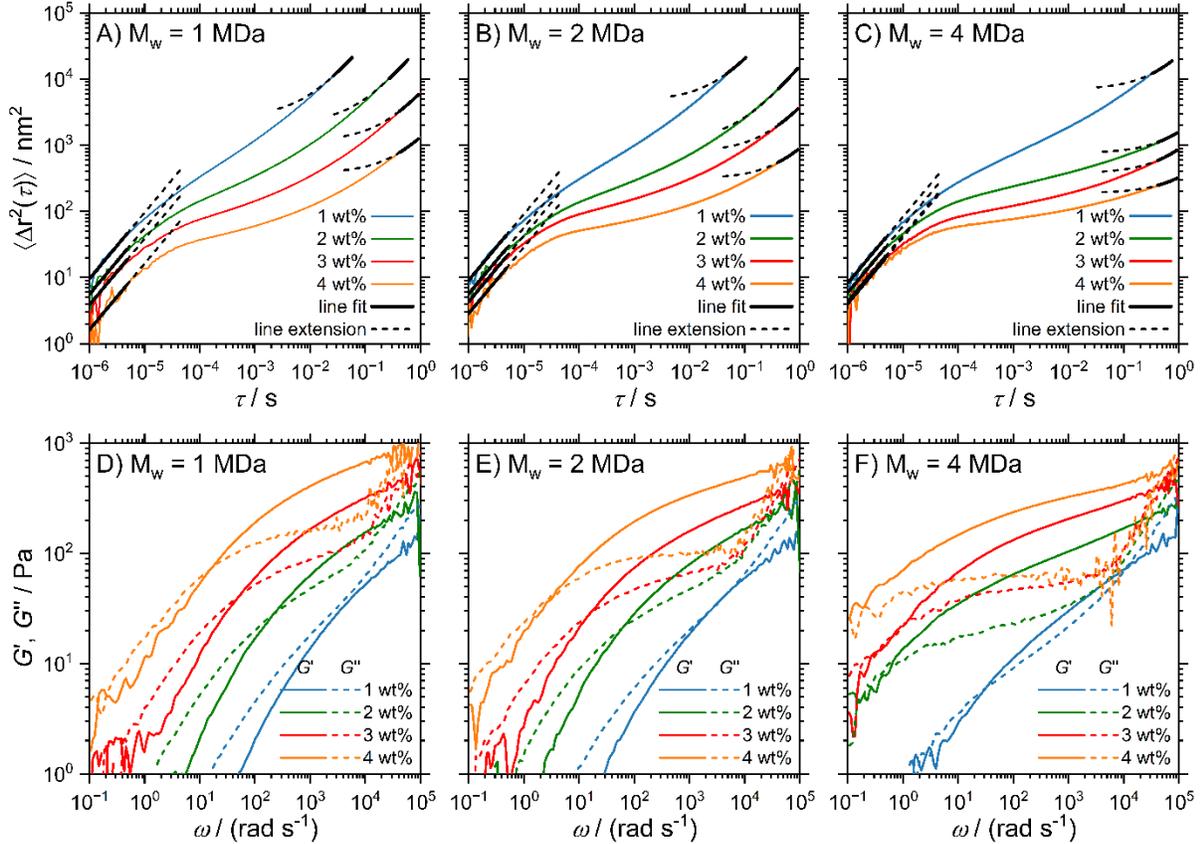

**Figure 2.** (A-C) Mean-squared displacements $\langle\Delta r^2(\tau)\rangle$ and (D-F) storage ($G'$) and loss moduli ($G''$) determined using DLS microrheology on PEO solutions containing PS-109 tracer particles. The full black lines in (A-C) indicate asymptotic linear fits, which have been extended by one

decade (broken black lines). The value of the constant in the long-time linear fits is substantial, explaining their curvature in the log-log plots.

The MSD is related to the frequency-dependent storage and loss moduli by the generalized Stokes-Einstein relation (GSER) [5,6,25,43]

$$G'(\omega) = |G^*(\omega)| \cos[\pi\alpha(\omega)/2] \, ,$$

$$G''(\omega) = |G^*(\omega)| \sin[\pi\alpha(\omega)/2] \, , \qquad (1)$$

with

$$|G^*(\omega)| = \frac{k_B T}{\pi a \langle \Delta r^2(1/\omega) \rangle \Gamma[1+\alpha(\omega)]} \, , \qquad (2)$$

where $k_B$ is the Boltzmann constant, $T$ the temperature, $a$ the hydrodynamic tracer-particle radius and $\omega$ the angular frequency. Here, $\Gamma(z) = \int_0^\infty x^{z-1} \mathrm{e}^{-x} \mathrm{d}x$ denotes the Gamma function. The MSDs are expressed as power laws with frequency-dependent exponent $\alpha(\omega)$ and converted into viscoelastic moduli (see SI Sections S5 and S6).[25,43] The results for the PS-109 samples are shown in Figure 2D-F.

Neglecting finite particle mass, in a purely viscous liquid, the particle MSD is linear in time. Particles trapped in a purely elastic solid never leave their initial position, so the MSD is constant. For viscoelastic hydrogels, three consecutive scaling regimes occur. At very short times, polymers do not influence the particle dynamics, which is determined only by the solvent viscosity,[44] $\langle \Delta r^2(\tau) \rangle = 6 D_{\text{solv}} \tau$, where $D_{\text{solv}}$ is the particle diffusion coefficient in pure solvent. We determine $D_{\text{solv}}$ from a fit according to $\langle \Delta r^2(\tau) \rangle = 6 D_{\text{solv}} \tau$ of the short-time MSD (see SI Section S7), for $10^{-6} < \tau < 5 \times 10^{-6}$ s. The solvent viscosity $\eta_{\text{solv}}$ follows from the Stokes-Einstein equation $D_{\text{solv}} = k_B T / (6\pi\eta_{\text{solv}} a)$. At intermediate times, the particles exhibit subdiffusion, $\langle \Delta r^2(\tau) \rangle \sim \tau^\alpha$ with $0 <$ $\alpha < 1$, reflecting hydrogel viscoelasticity. At very long times, the MSD becomes diffusive again, $\langle \Delta r^2(\tau) \rangle = 6D_{\text{micro}}\tau + b$, where $D_{\text{micro}} = k_B T / (6\pi\eta_{\text{micro}}a)$ characterizes the linear hydrogel viscosity $\eta_{\text{micro}}$ and $b$ is a constant shift.[18,45] Three measurements were performed per chain length and concentration, one for each tracer-particle radius $a$. Since no significant differences were found for varying $a$, the three values of $\eta_{\text{solv}}$ and $\eta_{\text{micro}}$ were averaged and are shown in Figure 3A.

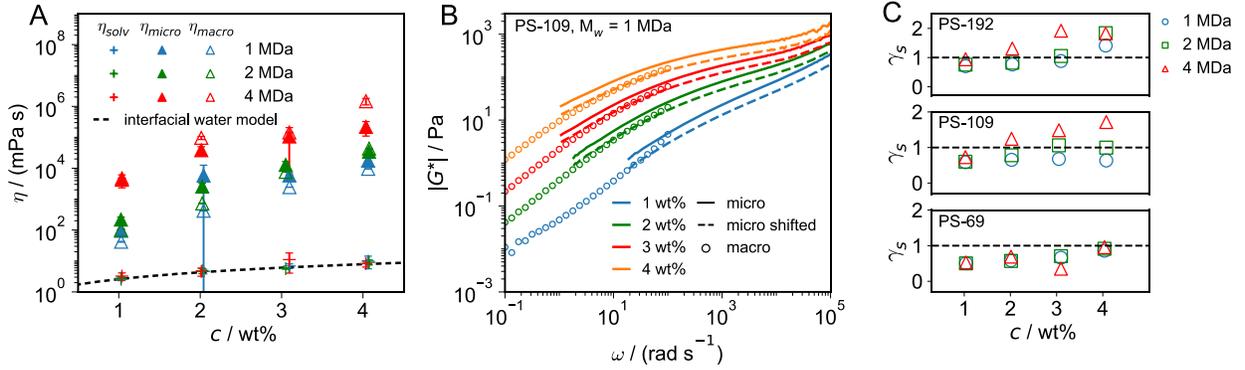

**Figure 3**. (A) Viscosities of the solvent and the hydrogel, $\eta_{\text{solv}}$ and $\eta_{\text{micro}}$, determined from linear fits of the short- and long-time behavior of the MSDs extracted from microrheology in Figure 2, compared to $\eta_{\text{macro}}$, determined from macrorheological steady-shear experiments. The broken line indicates the effective solvent viscosity of a polymer solution according to eq 4, which accounts for the increased viscosity of interfacial water surrounding the polymers. (B) Comparison of the viscoelastic moduli $|G^*| = \sqrt{(G')^2 + (G'')^2}$ from macro- and microrheological measurements for PS-109 tracer particles in 1 MDa PEO solutions (for the other datasets see SI Section S8). Circles denote macrorheology and solid lines microrheology results. Broken lines denote the microrheological data that is shifted by a factor $\gamma_s$ to match the macrorheology data (see SI Section S8). (C) Shift factor $\gamma_s$ for different tracer-particle sizes and PEO molecular weights (○: 1 MDa, □: 2 MDa, △: 4 MDa) as a function of PEO concentration. The black horizontal line denotes $\gamma_s = 1$, i.e., perfect agreement between macro- and microrheology.

The extracted solvent viscosity $\eta_{\text{solv}}$ in Figure 3A increases with polymer concentration $c$ but, expectedly, is independent of the chain length. The values for $\eta_{\text{solv}}$ range from 2 to 15 mPa s and are thus significantly larger than the viscosity of pure water at 25°C, which is $\eta_{\text{w}} = 0.89$ mPa s. In molecular dynamics simulations it was shown that the interfacial water layer at a polar surface exhibits a significantly increased water viscosity.[46] The thickness of that interfacial layer was obtained as $d = 0.4$ nm. To explain the increase of $\eta_{\text{solv}}$ with $c$, we regard each PEO polymer as being surrounded by an interfacial water layer with increased viscosity $\eta_i$. We model the hydrated polymers as cylinders with radius $R_{\text{cyl}} = (R_{\text{PEO}} + d)$, where $R_{\text{PEO}} = 0.229$ nm is the radius of a stretched PEO chain, estimated from the density of a PEO melt (see SI Section S9). The volume fraction of hydrated polymers is then given by

$$\phi_{\text{i}} = \frac{\pi(R_{\text{PEO}}+d)^2 c\, a_0\, \rho_{\text{solv}}\, N_{\text{A}}}{M_{\text{mono}}(100-c)},\qquad (3)$$

where $c$ is the polymer mass percentage, $a_0 = 0.356$ nm is the PEO monomer length,[47] $\rho_{\text{solv}}$ is the water mass density, $N_{\text{A}}$ is Avogadro's constant and $M_{\text{mono}} = 44.05$ g/mol is the molar mass of a PEO monomer. From $\phi_{\text{i}}$, the overall solvent viscosity follows from a simple geometric model (see SI Section S10) as

$$\eta_{\text{solv}} = \phi_{\text{i}}\eta_i + (1 - \phi_{\text{i}})\eta_w,\qquad (4)$$

where $\eta_i$ and $\eta_w$ are the viscosities of interfacial and bulk water, respectively. Using $\eta_w = 0.89$ mPa s and $d = 0.4$ nm, the fit of eq 4 to our experimental data (broken line in Figure 3A) yields $\eta_i = (27.17 \pm 0.74)$ mPa s, in good agreement with the simulation results.[46] We thus conclude that the increase of the solvent viscosity from microrheology can be well explained by the increased viscosity of interfacial water layers around PEO.

Additionally, the hydrogel viscosity was extracted from non-oscillatory macrorheological measurements at steady shear rate $\dot{\gamma}$ by fits to the non-linear Cross model (see SI Section S11).

Since $\eta_{\text{macro}}$ is the limiting value for zero shear rate, it is the linear-response viscosity that can be compared to $\eta_{\text{micro}}$ from microrheology. As evidenced in Figure 3A, $\eta_{\text{macro}}$ and $\eta_{\text{micro}}$ are comparable, but systematic shifts are observed, as will be discussed and explained in detail below.

**Comparison between macro- and microrheology.** In Figure 3B we compare the absolute values of the viscoelastic modulus $|G^*| = \sqrt{(G')^2 + (G'')^2}$ from microrheology and macrorheology for tracer particle PS-109 and polymer weight $M_w = 1$ MDa. Deviations are quantified by a frequency-independent shift factor $\gamma_s$ according to $|G^*_{\text{micro,shifted}}| = \gamma_s |G^*_{\text{micro}}|$ (see SI Section S8), where a value $\gamma_s = 1$ indicates the validity of the GSER. The shifted $|G^*_{\text{micro,shifted}}|$, shown in Figure 3B as broken lines, perfectly agree with the macrorheological data. Some discrepancies are observed for the samples with longer polymers, presumably due to inaccuracies of macrorheological measurements at high frequencies due to inertial effects (see SI Section S3) as well as long polymeric relaxation times. In Figure 3B, $\gamma_s$ is demonstrated to systematically increase with polymer concentration, while there is a much weaker and less clear dependence on tracer-particle size and chain length.

To investigate the mechanism behind the discrepancies between macro- and microrheology and the salient dependence of the shift factor $\gamma_s$ on polymer concentration, we derive a generalized GSER from the transient Stokes equation around a sphere of radius $a$ that includes slip on the sphere surface and compressibility in the embedding fluid. The transient Stokes equation includes a general frequency-dependent viscosity and thus correctly accounts for the fluid viscoelasticity. As detailed in SI Section S12, we find no significant effects due to the finite-sphere radius for $a$ below 10 μm in the experimental frequency range of $10^{-1} < \omega < 10^5$ rad/s. Also, finite slip always decreases the particle friction, in contrast to the deviation between macro- and microrheology in Figure 3B, which for some experiments suggests strong enhancement of particle friction. Thus, the GSER in eq 2, which neglects finite sphere radius, compressibility and slip effects, is for the employed particle radii and particle types an accurate approximation of the exact solution of the transient Stokes equation derived in SI Section S12.

The GSER eq 2 furthermore assumes a homogeneous viscoelastic medium and thus neglects the hydrogel structuring on the scale of the mesh size $\xi$.[39,48,49] For particle radii $a \gg \xi$ this assumption is warranted,[16,18,50] for smaller particles deviations are expected.[51] Since the mesh size is experimentally only indirectly accessible,[52] it is often estimated by the polymer correlation length, $\xi_{SANS}$, as obtained from scattering experiments.[53–56] Depending on the PEO concentration, values of $\xi_{SANS} \approx 2 - 8$ nm were found in our SANS measurements (see SI Section S13). These lengths favorably compare to the simple cubic-lattice estimate $\xi_{cubic} = \left(\frac{3}{a_0 \phi_m}\right)^{1/2}$, where $\phi_m$ is the monomeric number density (see SI Section S14). We obtain $\xi_{cubic} = 3.9$ nm for 4 wt% PEO and $\xi_{cubic} = 7.9$ nm for 1 wt% PEO, in good agreement with our SANS measurements. Since the estimated mesh sizes are much smaller than the tracer-particle radii used, which range from diameters of 69 nm to 192 nm, we conclude that the hydrogels are homogeneous on the tracer-particle size and deviations between macro- and microrheology cannot plausibly be explained by inhomogeneity effects in the bulk hydrogel.

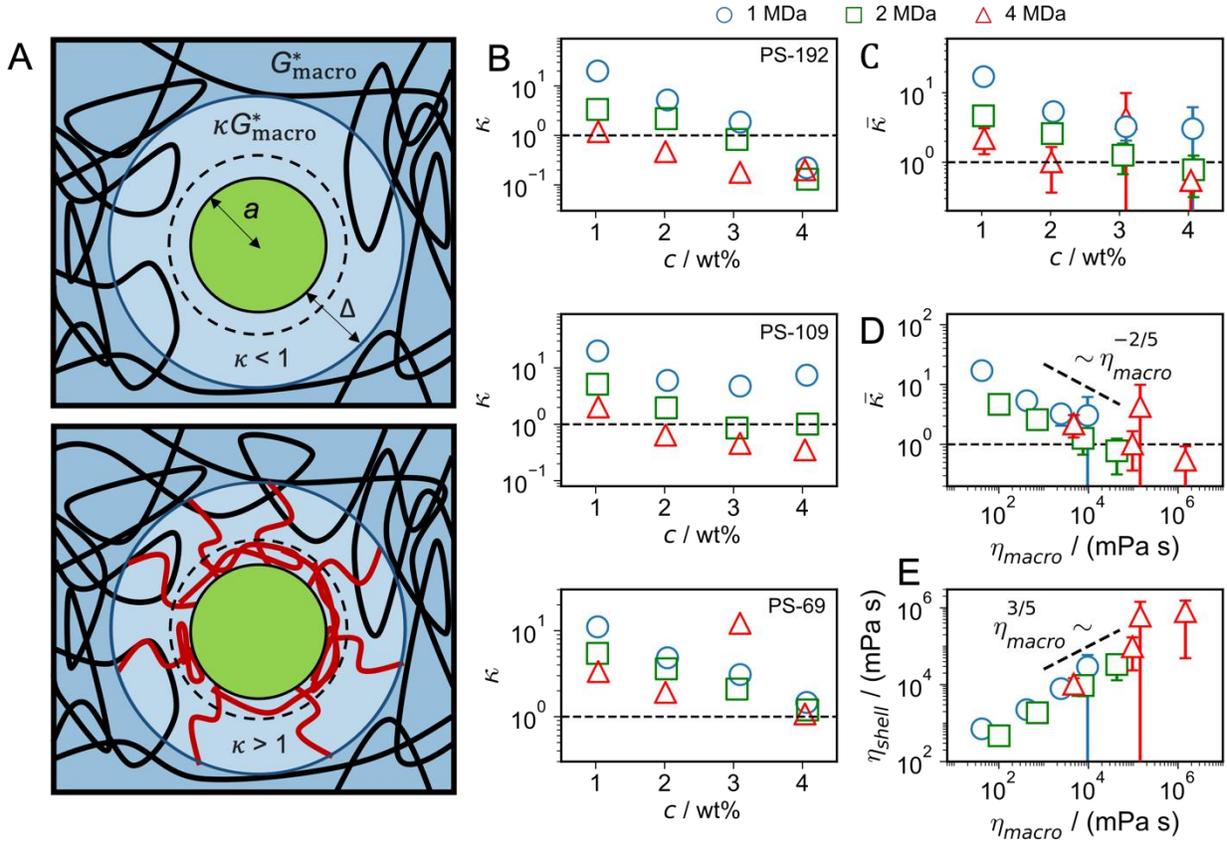

**Figure 4.** (A) Sketch of a tracer particle in a PEO hydrogel. Particle-PEO interactions produce a depletion (top) or adsorption layer (bottom), indicated by broken circles, within which the PEO density differs from bulk. Consequently, the viscoelastic polymer response $G_{\text{shell}}^*(\omega)$ deviates from the bulk spectrum $G_{\text{macro}}^*(\omega)$ within a shell of thickness $\Delta$ (indicated by solid circles). The viscoelastic shell thickness $\Delta$ in the adsorption case (bottom) is dominated by dangling adsorbed chains (shown in red) and therefore is larger than the adsorption layer thickness. (B) Ratio of shell and bulk viscoelasticity $\kappa = G_{\text{shell}}^*(\omega)/G_{\text{macro}}^*(\omega)$, which follows from the shift factor $\gamma_s$ in Figure 3C, as a function of polymer concentration for different tracer-particle radii and PEO molecular weights. Fit errors are much smaller than the symbol size. (C, D) Interfacial viscoelastic enhancement factor averaged over the results for different tracer radii in Figure 4B, $\bar{\kappa}$, plotted as a function of (C) the polymer concentration and (D) the bulk viscosity $\eta_{\text{macro}}$. Vertical bars indicate

the standard deviation of the average over the tracer particle radii and are only shown if larger than the symbol size. (E) Interfacial shell viscosity $\eta_{\text{shell}} = \bar{\kappa}\eta_{\text{macro}}$ in dependence of bulk viscosity $\eta_{\text{macro}}$. Power laws are added as guides to the eye.

We therefore consider an alternative mechanism for the GSER violation. The GSER assumes the hydrogel around the tracer particles to be entirely described by the bulk modulus $G^*_{\text{macro}}(\omega)$, but due to perturbations of the hydrogel around the particles, a shell with a thickness $\Delta$ and a different modulus $G^*_{\text{shell}}(\omega)$ will in general be present around tracer particles. As illustrated in Figure 4A, the shell within which the modulus differs from bulk will in general have a different thickness $\Delta$ than the layer within which the polymer density differs from the bulk value, indicated by a broken circle.

The particle-polymer interactions can be repulsive or attractive and induce depletion[56–62] (upper scheme) or adsorption layers[37,63–65] (lower scheme), respectively. For depletion one expects a shell with a reduced modulus, which would lead to a finite slip, for adsorption one expects an increased shell viscoelastic modulus. To reduce the number of free variables in our shell model, we assume that the shell modulus is related to $G^*_{\text{macro}}(\omega)$ by a frequency-independent factor according to $G^*_{\text{shell}}(\omega) = \kappa G^*_{\text{macro}}(\omega)$. The modified generalized Stokes-Einstein relation for such a shell model has been derived from the Stokes equation and reads[39]

$$|G^*(\omega)| = \frac{k_{\text{B}}T}{\pi a \langle \Delta r^2(1/\omega)\rangle \Gamma[1+\alpha(\omega)]}\gamma_s(\Delta, \kappa), \qquad (5)$$

the explicit form of the correction factor $\gamma_s(\Delta, \kappa)$ is given in SI Section S15. If $\Delta = 0$ or $\kappa = 1$ one has $\gamma_s(\Delta, \kappa)=1$ and eq 5 converges to eq 2. Alternatively, our data could be rationalized by a modified effective tracer radius [65–68], but we argue that a decreased shell viscoelastic response is a more physical model than a decreased effective tracer radius (see SI Section S16). The parameters $\Delta$ and $\kappa$ cannot be simultaneously determined from the experimentally measured $\gamma_s$ values in Figure 3C, as explained in SI Section S17. By analysis of the deviation between macro- and microrheological data, we find that the shell thickness $\Delta$ is linearly related to the polymer end-to-end distance $R_e^{\text{ideal}}$, which suggests that the viscoelastic perturbation in the interfacial shell is transmitted by polymers that adsorb to the particle surface and dangle into solution, in line with literature results for the hydrodynamic radius of adsorbed polymer layers.[37,69–78] We therefore take $\Delta$ proportional to $R_e^{\text{ideal}}$ and determine $\kappa$ by inversion of $\gamma_s(\Delta, \kappa)$ for each experiment. The proportionality constant between $\Delta$ and $R_e^{\text{ideal}}$ is assumed identical for all systems and chosen as the minimal value that describes all experimental $\gamma_s$ values, see SI Section S17 for details. We obtain $\Delta = \frac{3}{5} R_e^{\text{ideal}}$, where the values of $R_e^{\text{ideal}}$ are given in SI Section S1.

In Figure 4B, the results for $\kappa$ are shown to range between 0.1 and 20 and to generally decrease with increasing polymer concentration with a smaller dependence on particle size (see SI Section S18). We therefore average over different particle radii, the resulting average $\bar{\kappa}$ in Figure 4C is seen to decrease with concentration and reaches $\bar{\kappa} \approx 1$ for high concentration. This means that the effect of the adsorbed polymer chains on the rescaled viscoelastic modulus in the interfacial shell diminishes with increasing bulk polymer concentration, in line with the fact that the relative increase of polymer concentration in the adsorbed surface layer also decreases with increasing bulk polymer concentration.[56,79–82] Also, $\bar{\kappa}$ in Figure 4C decreases with increasing polymer chain length, which is plausible since the slowing down of the shell dynamics due to adsorbed polymer chains becomes less important compared to the slowing down due the hindered reptation as the polymer chains become longer.[37,83,84] To investigate the relation between the interfacial-shell and the bulk viscosity, we plot in Figure 4D the shell/bulk modulus ratio $\bar{\kappa}$ versus the bulk viscosity $\eta_{\text{macro}}$. In this scaling plot an approximate data collapse between different polymer chain lengths occurs and we see that the relative increase of the viscosity in the interfacial shell decreases significantly and almost universally with bulk viscosity $\eta_{\mathrm{macro}}$. Clearly, we expect the relation between $\bar{\kappa}$ and $\eta_{\mathrm{macro}}$ to depend on the surface material, which we did not vary in the current study. The added straight line is merely meant as guide to the eye and not as proof of a power law. In Figure 4E we show the interfacial shell viscosity $\eta_{\mathrm{shell}} = \bar{\kappa}\eta_{\mathrm{macro}}$ as a function of the hydrogel bulk viscosity $\eta_{\mathrm{macro}}$, which demonstrates that the shell viscosity increases dramatically with rising bulk viscosity. Although the experimental data is scarce at the highest bulk viscosities, we presume that the shell viscosity $\eta_{\mathrm{shell}}$ increases linearly with the bulk viscosity $\eta_{\mathrm{macro}}$ for $\eta_{\mathrm{macro}} > 10^5$ mPa s, so that $\eta_{\mathrm{shell}}$ is never smaller than $\eta_{\mathrm{macro}}$. This reflects that the polymers adsorb onto the particles and therefore the polymer density is increased close to the particle surface.

**Conclusions**. We demonstrate that the GSER is an accurate theoretical model to extract viscoelastic properties from microrheology and that observed deviations between macro- and microrheology data can be explained by interfacial effects in a shell around the tracer particles. The shell thickness is proportional to the polymer end-to-end distance and thus significantly larger than the structural adsorption layer thickness measured in scattering experiments,[64,85,86] which reflects the importance of chain ends that dangle into the solution for the rheological properties around the particles.[37,69,70] This not only reconciles micro- and microrheological measurements but also gives insights into the interfacial viscoelastic behavior of hydrogels and polymer solutions. Our methods are general and can be applied to more complex viscoelastic fluids and particles to investigate their interfacial rheological properties. In the future, it would be desirable to extract the detailed frequency-dependent shell viscoelastic modulus; for this, experiments over extended frequencies would have to be performed.

ASSOCIATED CONTENT

**Supporting Information**

The Supporting Information is attached as separated file.

S1: Scaling relations of PEO solutions. S2: Materials and sample preparation. S3: Macrorheology experimental details. S4: Fractional Maxwell model. S5: Microrheology experimental details. S6: Determination of the frequency-dependent power law exponent $\alpha(\omega)$. S7: Determination of $\eta_{\text{solv}}$ and $\eta_{\text{micro}}$ from microrheology. S8: Determination of the shift factor. S9: Estimation of molecular PEO radius. S10: Model for effective solvent viscosity. S11: Steady-shear experiments. S12: Transient and compressibility effects in PEO solutions. S13: Small-angle neutron scattering of PEO solutions. S14: Mesh size of a cubic polymer network. S15: Derivation of the shell-model GSER. S16: Effective tracer radii from experimental shifts. S17: Adjusting the dynamic moduli data using the shell model. S18: Dependence of shell modulus on tracer size. S19: Specification of particles. S20: Amplitude sweep results.

AUTHOR INFORMATION

**Corresponding Author**

*E-mail: rnetz@physik.fu-berlin.de

**Author Contributions**

Conceptualization: MG and RN. Data Curation: RS and HK. Formal Analysis: RS and HK. Investigation: RS and HK. Methodology: HK and RN. Software: HK and RN. Writing/Original Draft Preparation: RS, HK, MG and RN. Writing/Review & Editing: RS, HK, MG, RN.

**Funding**

This study and the authors were funded by the Deutsche Forschungsgemeinschaft (DFG, German Research Foundation)–SFB 1449–431232613 project A02.

**Data Availability**

The data that supports the findings of this study are available from the corresponding author upon reasonable request.

**Conflict of Interests**

The authors have no conflicts to disclose.

ACKNOWLEDGMENT

We thank the *Deutsche Forschungsgemeinschaft (DFG)* for funding this work as part of the collaborative research center CRC 1449 on 'Dynamic Hydrogels at Biointerfaces'. Further, RS would like to acknowledge the *Fonds der chemischen Industrie* for financial support. We gratefully acknowledge the ISIS Pulsed Neutron and Muon Source (STFC Rutherford Appleton Laboratory, Didcot, U.K., experiment RB2220343 [87]) for granting neutron beamtime for this project.

ABBREVIATIONS

PEO, polyethylene oxide; GSER, generalized Stokes-Einstein relation; DLS, dynamic light scattering; PS, polystyrene, MSD, mean-squared displacement; SI, Supporting Information

For Table of Contents Only

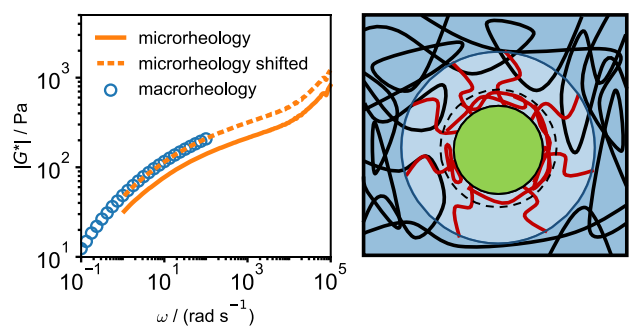

# SUPPORTING INFORMATION for:

# Nanoscopic Interfacial Hydrogel Viscoelasticity Revealed from Comparison of Macroscopic and Microscopic Rheology

*Robert F. Schmidt [a), Henrik Kiefer [b), Robert Dalgliesh [c), Michael Gradzielski [a), Roland R. Netz[*,b)*

*[a) Stranski-Laboratorium für Physikalische und Theoretische Chemie, Technische Universität Berlin, Strasse des 17. Juni 124, 10623 Berlin, Germany*

*[b) Fachbereich Physik, Freie Universität Berlin, Arnimallee 14, 14195 Berlin, Germany*

*[c) STFC, ISIS, Rutherford Appleton Laboratory, Chilton, Oxfordshire OX11 0QX, United Kingdom*

\* E-mail: rnetz@physik.fu-berlin.de

## S1: Scaling relations of PEO solutions

Solutions of linear polymers are viscoelastic in the semi-dilute regime for concentrations above the overlap concentration $c^*$, where individual polymer coils begin to overlap.[1–3] Based on the radius of gyration $R_g$ as a measure for the effective extension of dilute coils and the molecular weight $M_w$ of the polymer, we approximate the overlap concentration $c^*$ by[4]

$$c^* = \frac{3M_w}{4N_A \pi R_g^3}, \qquad (S1)$$

where $N_A$ is Avogadro's constant. Devenand and Selser[5] introduced an empirical relationship for the radius of gyration of dilute poly(ethylene oxide) (PEO) in water, which is a good solvent for PEO, as a function of molecular weight

$$R_g/\text{Å} = 0.215(M_w/(\text{g mol}^{-1}))^{0.583\pm0.031} \quad . \qquad (S2)$$

Using eqs S1 and S2, we calculate $R_g$ and $c^*$ for the PEO molecular weights used in this work. In the semi-dilute regime, i.e., for concentrations above $c^*$, repulsive interactions between monomers in a single polymer chain are screened by the presence of other polymer chains, and therefore, beyond the mesh size, chains can be considered as ideal.[6] The average end-to-end distance $R_e^{\text{ideal}}$ of an ideal chain is given by $R_e^{\text{ideal}} = \sqrt{ba_0N}$,[6] where $N$ is the number of monomers. For the Kuhn and monomer lengths of PEO we use $b = 0.68$ nm and $a_0 = 0.356$ nm, respectively.[7] Furthermore, we can readily calculate the radius of gyration of an ideal chain $R_g^{\text{ideal}} = \sqrt{N}a_0/\sqrt{6}$. The results for $R_g$, $R_e^{\text{ideal}}$, $R_g^{\text{ideal}}$ and $c^*$ are shown in Table S1. $R_g$ determined from the empirical relationship in eq S2 lies between $R_e^{\text{ideal}}$ and $R_g^{\text{ideal}}$, except for 4 MDa. All concentrations of PEO used in this work are semi-dilute since their polymer concentrations are above the respective overlap concentration $c^*$.

**Table S1.** Radius of gyration $R_g$ in dilute solution, ideal end-to-end distance $R_e^{ideal}$ and ideal radius of gyration $R_g^{ideal}$, relevant for semi-dilute solutions, and overlap concentration c* for PEO with three different molecular weights $M_w$.

| $M_w$ / (g mol$^{-1}$) | $N$ | $R_g$ / nm | $R_e^{ideal}$ / nm | $R_g^{ideal}$ / nm | $c^*$ / %w/v |
|---|---|---|---|---|---|
| 1 x 10$^6$ | 2.27 x 10$^4$ | 68 | 74 | 22 | 0.13 |
| 2 x 10$^6$ | 4.54 x 10$^4$ | 101 | 105 | 31 | 0.08 |
| 4 x 10$^6$ | 9.08 x 10$^4$ | 152 | 148 | 44 | 0.05 |

## S2: Materials and sample preparation

The hydrogels used in this work are aqueous solutions of PEO with average molecular weights $M_w$ of 1, 2 and 4 MDa obtained from Sigma-Aldrich. For the dynamic light scattering (DLS) microrheology measurements, we used polystyrene (PS) tracer particles with hydrodynamic diameters of 68.8, 109.3 and 192.0 nm (referred to as PS-69, PS-109 and PS-192) and polydispersity indices of 3.93%, 1.86% and 3.10%, respectively, as determined by DLS in Supporting Information Section S19. All PS particles were obtained from Polysciences as 2.6–2.7 %w/v aqueous suspensions and contain a slight anionic charge from sulfate ester to prevent agglomeration.

All samples were prepared by adding the appropriate mass of polymer into a cylindrical glass vial. The required volume of Milli-Q water containing the respective tracer particles was then added using an Eppendorf pipette. The precise masses (± 1 mg) of polymer and particle solution were determined using an analytical balance. All concentrations are given as weight percentages %w/w, thereby being independent of temperature. The tracer particle concentration was 0.003 %w/v, 0.01 %w/v and 0.04 %w/v for PS-192, PS-109 and PS-69, respectively. Prior to adding the particle solutions to the polymer, they were sonicated for 15 min to break up possible particle agglomerations. After adding all components, the samples were stirred at ambient temperature using a magnetic stirrer until they appeared fully homogenous for at least 24 h. Very highly viscous samples required up to 3 days of stirring until being fully homogenized. Samples used for small-angle neutron scattering (SANS) experiments (see Supporting Information Section S13) were prepared with $D_2O$ instead of $H_2O$ and without any tracer particles since they would otherwise dominate the scattering spectra. The $D_2O$ was filtered with a 0.2 µm cellulose acetate filter prior to use. The $D_2O$ samples were prepared to attain the same weight per volume percentage %w/v as their $H_2O$ counterparts at 25 °C.

## S3: Macrorheology experimental details

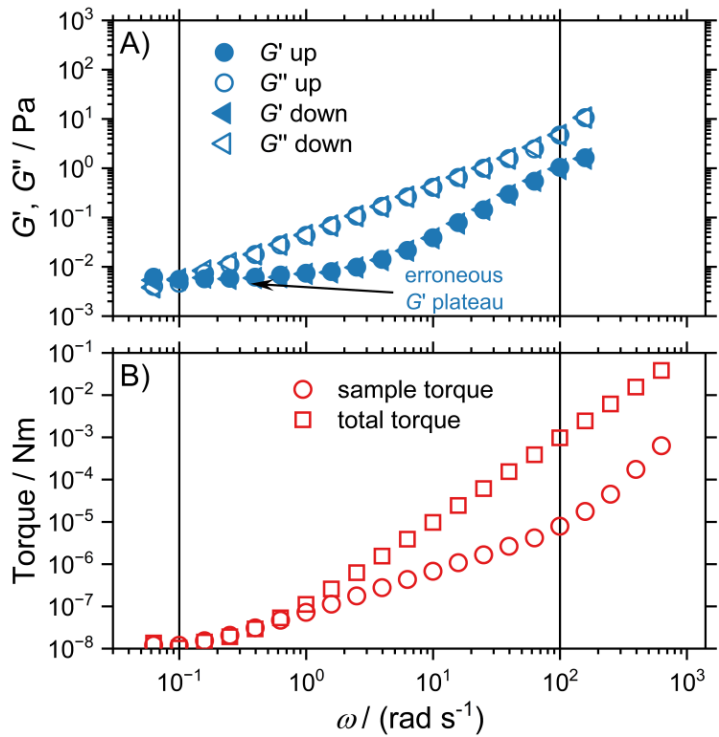

**Figure S1**. Illustration of macrorheological experimental limitations for the low-viscous sample with 1 MDa 1 wt% PEO. A) Storage ($G'$) and loss ($G''$) moduli from up (circles) and down (triangles) frequency sweeps demonstrate perfect agreement. The storage modulus exhibits a spurious plateau at low frequencies. B) Comparison of total and sample torque. The vertical black lines indicate the frequency range used in the main text for the comparison with microrheological measurements.

All macrorheological measurements were performed on an MCR 502 WESP temperature-controlled rheometer from Anton Paar (Graz, Austria) in strain-imposed mode, which has a combined motor transducer (CMT) design. The macrorheology experiments were performed on the same particle-containing samples that were used in microrheology. A cone-and-plate measuring system with a diameter of 50 mm and a cone angle of 1 was used. After the sample (volume = 650 μL) was loaded onto the lower plate, the cone was lowered slowly until the final gap width of 101 μm was reached. The temperature of the lower plate as well as the cone were set to 25°C using Peltier elements. Sample, plate and cone are enclosed by a hood to ensure a homogeneous temperature around the sample. After the target temperature of 25°C was reached (time required ≈ 1 min), the sample was left to equilibrate for 3 min. For the PS-192 containing samples, two different types of measurements were performed. In a first oscillatory measurement (amplitude sweep, duration ≈ 5 min), the angular frequency $\omega$ was kept constant at 6.28 rad/s, and the strain amplitude $\gamma_0$ was varied from 0.1 to 20% to determine the linear viscoelastic regime (LVE). A strain amplitude of $\gamma_0 = 5\%$ was chosen for all subsequent frequency sweeps. The amplitude sweep results are shown in Supporting Information Section S20. Secondly, in a frequency sweep measurement, the strain amplitude was kept fixed at $\gamma_0 = 5\%$, while the angular frequency $\omega$ was increased from 0.1 and 100 rad/s (up-sweep, duration ≈ 16 min). Additionally, the frequency was afterwards decreased (down-sweep, duration ≈ 16 min) to check for hysteresis effects. The data shown in the main text represent the up-sweep only (total time until end of up-sweep ≈ 25 min). Since the up- and down-sweeps superimpose very well (see Figure S1), we can neglect hysteresis effects. We can also rule out evaporation of significant amounts of solvent, which would lead to differences in the up- and down sweeps. For the PS-109 and PS-69 containing samples, no amplitude sweep was performed, bringing the total time until the end of the up-sweep to 20 min. For measurements of multiple samples of the same concentration (e.g., three measurements of 2000 kDa, 2 wt% with PS-192, PS-109 and PS-69 particles), the exact values of $G$' and $G$'' can differ slightly due to small differences in concentration and/or small differences in the sample volume loaded onto the rheometer plate. This is especially noticeable when the overall measurement signal is low, i.e., for low viscous samples and at low frequencies.

Lastly, for the samples containing PS-109 and PS-69 particles, steady-shear experiments were performed in addition (see Supporting Information Section S11). Here, the shear rate $\dot{\gamma}$ was varied between 0.1 and 100 s$^{-1}$, first in increasing and then in decreasing order. All samples showed shear-thinning behavior at higher shear rates.

For samples with very low viscoelasticity, challenges arise due to instrumental limitations. This is demonstrated in Figure S1 for a 1 wt% solution of 1 MDa PEO, which is the lowest viscous sample studied in this work. For the purpose of this demonstration, the frequency range was extended to frequencies below and above the range used for all remaining samples in the main text (as indicated by the vertical black lines in Figure S1). At low frequencies, $G$' approaches a frequency-independent plateau, which is very different from the terminal power-law scaling of $\sim\omega^2$ expected according to the Maxwell model. This plateau is an artefact and has recently been attributed to phase-angle uncertainties.[8,9] It occurs mostly for fluid samples with a low torque signal. As indeed shown in Figure S1B, the sample torque reaches very low values of $\sim10^{-8}$ Nm at low frequencies.

At very high frequencies, instrument inertia can lead to artefacts in the data. The measured total torque consists of the sample torque and the torque necessary to accelerate the moving components of the instrument, where the latter is automatically subtracted. When the total torque becomes significantly higher than the sample torque by a factor of around 2 orders of magnitude, artefacts can occur.[10] This is usually the case for low viscous samples at high frequencies, such as the PEO solution shown in Figure S1. For the three highest frequencies, the rheometer displays an error message indicating that $G$' and $G$'' can no longer be accurately determined, which is why they are not shown in Figure S1A. In the frequency range used in the main text, which is indicated by black vertical lines in Figure S1, the instrument inertia effect is small but might still be noticeable for the low-viscous samples at the highest frequencies.

**S4: Fitting of macrorheology data**

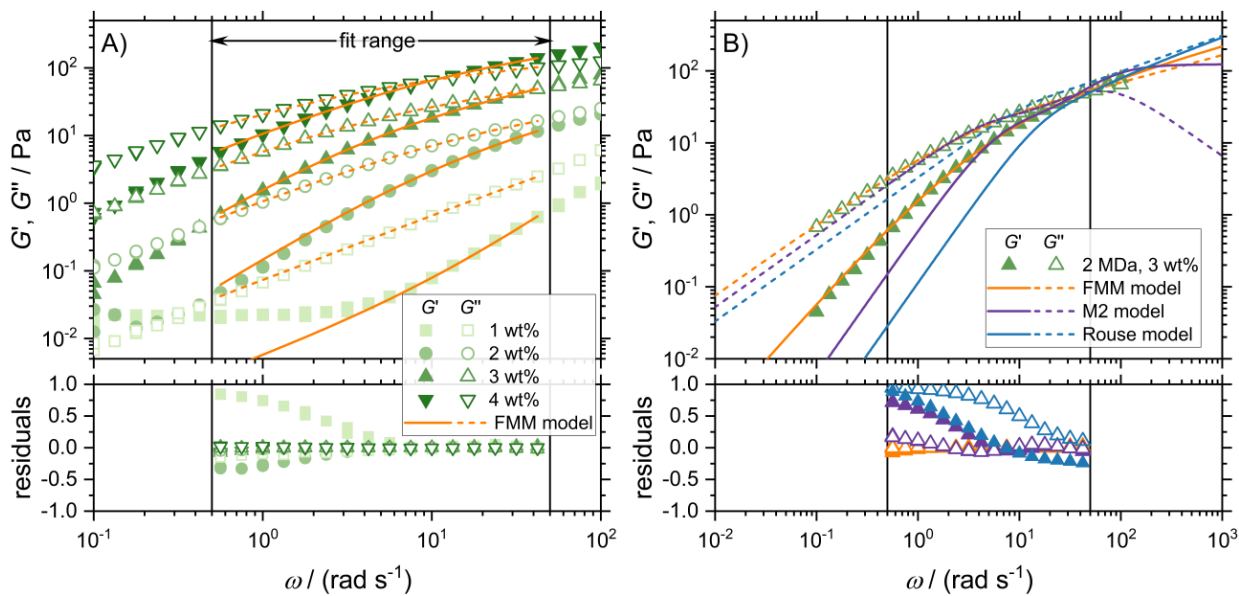

**Figure S2**. A) Frequency sweep data of the 2 MDa PEO solutions fitted with the fractional Maxwell model (FMM). The fit yields a very good description over a wide range of frequencies. B) The FMM yields much better fit results than a simple two-mode Maxwell model (M2) or the Rouse model.

Classical viscoelastic models are composed of arrangements of elastic springs and viscous dashpots. In contrast, fractional viscoelastic models contain elements called *spring-pots* which employ fractional derivatives of time and have properties intermediate between an elastic spring and a viscous dashpot.[11] Instead of a sum of exponentials, they predict power law behavior for the time- and frequency-dependent material functions. The constitutive equation for a spring-pot is given by $\sigma(t) = \eta_\alpha \, d^\alpha \gamma(t)/dt^\alpha$, where $\sigma$ is the shear stress, $\gamma$ is the shear strain, $\alpha$ is a fractional exponent between 0 and 1, and $\eta_\alpha$ is a material property with units Pa s$^\alpha$.[12,13] If $\alpha = 0$, the spring-pot turns into a regular spring, and if $\alpha = 1$, it turns into a regular dashpot. It has been shown that fractional models can be realized by arrangements of an infinite number of springs and dashpots as ladders, trees, or fractal structures.[13–17] One particular fractional model is the fractional Maxwell model (FMM), which connects two spring-pots in a series. The complex modulus $G^*(\omega)$ for the FMM is given by

$$G^*(\omega) = \frac{\eta_\alpha (i\omega)^\alpha \eta_\beta (i\omega)^\beta}{\eta_\alpha (i\omega)^\alpha + \eta_\beta (i\omega)^\beta}, \qquad (S3)$$

where, by definition, $\alpha > \beta$. The real ($G'$) and imaginary parts ($G''$) can be separated using that $i^n = \cos(n\pi/2) + i \sin(n\pi/2)$.[18] The real ($G'$) and imaginary ($G''$) parts of the complex modulus of the fractional Maxwell model (FMM) are given by

$$G'(\omega) = \frac{\eta_\alpha \omega^\alpha \eta_\beta \omega^\beta [A \cos((\alpha+\beta)\pi/2) + B \sin((\alpha+\beta)\pi/2)]}{A^2 + B^2},$$

$$G''(\omega) = \frac{\eta_\alpha \omega^\alpha \eta_\beta \omega^\beta [A \sin((\alpha+\beta)\pi/2) - B \cos((\alpha+\beta)\pi/2)]}{A^2 + B^2}, \qquad (S4)$$

where

$$A = \eta_\alpha \omega^\alpha \cos(\alpha\pi/2) + \eta_\beta \omega^\beta \cos(\beta\pi/2),$$

$$B = \eta_\alpha \omega^\alpha \sin(\alpha\pi/2) + \eta_\beta \omega^\beta \sin(\beta\pi/2). \qquad (S5)$$

Depending on the two fractional exponents $\alpha$ and $\beta$, the FMM can interpolate between a completely solid ($\alpha = \beta = 0$), a completely liquid ($\alpha = \beta = 1$), and a viscoelastic ($1 > \alpha > \beta > 0$) material. In the special case that $\alpha = 1$ and $\beta = 0$, the regular Maxwell model is retrieved. In Figure S2A, exemplary FMM fits are shown for the frequency sweeps of 2 MDa PEO solutions. The FMM also yields a better fit of the data than the generalized Maxwell model with two elements (M2) or the Rouse model, as shown in Figure S2B. The storage and loss moduli for the generalized Maxwell model are given by[19,20]

$$G'(\omega) = \sum_{i=i}^{N} g_i \frac{\omega^2 \tau_i^2}{1+\omega^2 \tau_i^2}$$

$$G''(\omega) = \sum_{i=i}^{N} g_i \frac{\omega \tau_i}{1+\omega^2 \tau_i^2} \qquad (S6)$$

where $N$ is the number of modes and $\tau_i$ and $g_i$ are the relaxation time and strength of mode i. The storage and loss moduli of the Rouse model are given by[21]

$$G'(\omega) = \frac{\rho RT}{M} \sum_{p=1}^{N} \frac{\omega^2 \tau_p^2}{1+\omega^2 \tau_p^2}$$

$$G''(\omega) = \frac{\rho RT}{M} \sum_{p=1}^{N} \frac{\omega \tau_p}{1+\omega^2 \tau_p^2} \qquad (S7)$$

where $\rho$ is the density, $R$ is the molar gas constant, $T$ is the temperature and $M$ is the molar mass of the polymer. There are $N$ relaxation modes in the Rouse model that are given by

$$\tau_p = \frac{6\eta_0 M}{\pi^2 p^2 \rho RT}, p = 1, 2, 3 \dots N \qquad (S8)$$

where $\eta_0$ is the zero-shear viscosity. The Rouse model is a special case of the generalized Maxwell model with $N$ equally weighted modes.

## S5: Microrheology experimental details

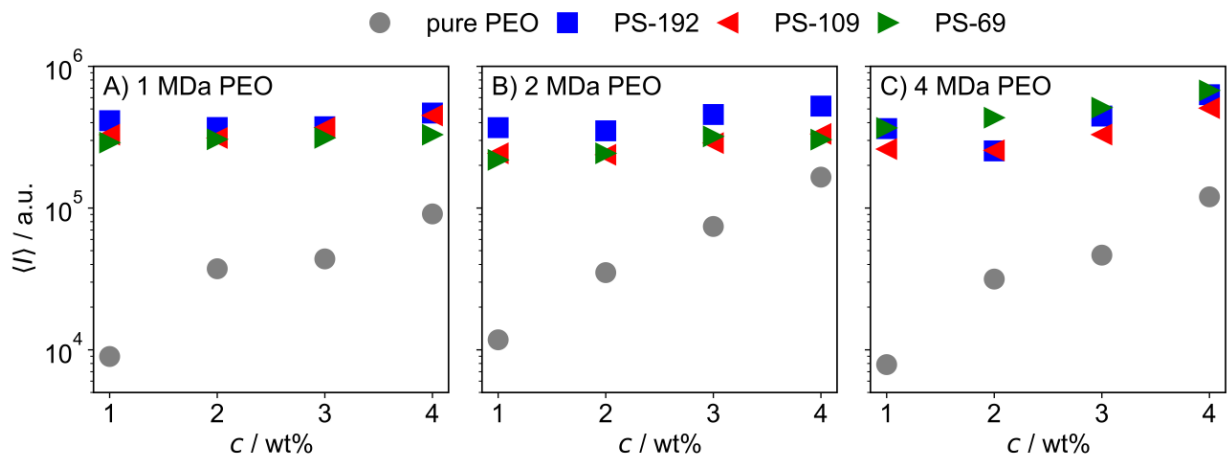

**Figure S3**. Static light scattering intensity $<I>$ for all PEO solutions both with and without added particles. The intensity of the particle-containing solutions is always significantly larger than of the pure solutions, ensuring that the measured correlation functions are dominated by the dynamics of the particles and not the polymers themselves.

Microrheology experiments were performed using dynamic light scattering (DLS) on a Litesizer 500 instrument from Anton Paar (Graz, Austria), equipped with a 40 mW semiconductor laser diode with a wavelength $\lambda = 658$ nm. The scattering angle $\theta$ was kept fixed at 175°. The modulus of the scattering vector is given by $q = 4\pi n/\lambda \; \sin(\theta/2)$, which yields a value of 0.025 nm$^{-1}$. Here, $n$ is the refractive index of the solvent. The measurements were done using a 3 x 3 mm low volume quartz cuvette, which requires a sample volume of around 50 μL. The temperature was kept constant at 25°C during the measurements. The measurement time was set to the maximum of 30 min and thus is much larger than the structural relaxation time $\tau_0 = 2\pi/\omega_0$, where $\omega_0$ is the viscoelastic crossover frequency (see Figure 1 in the main text). For all our samples, $\omega_0 > 0.1$ rad/s, meaning $1/\omega_0 < 63$ s, ensuring that structures relax in the time frame of the experiment. The static scattering intensity of the PEO solutions $<I>$ with and without particles is shown in Figure S3. Since the scattering intensity of the particle-containing samples is always significantly larger than of the pure PEO solutions, we can assume that the DLS signal is dominated by the dynamics of the particles and not the polymers themselves.

The data output consists of the intensity–intensity autocorrelation function

$$g^{(2)}(\tau) = \langle I(t)I(t+\tau)\rangle / \langle I(t)\rangle^2 \quad , \qquad \text{(S9)}$$

where $\tau$ is the lag time and $I$ is the intensity. The brackets $<\dots>$ denote a time average. For spatially coherent polarized light, the second-order correlation function $g^{(2)}(\tau)$ can be related to the first-order correlation function $g^{(1)}(\tau)$ using the Siegert relation [22,23]

$$g^{(2)}(\tau) = 1 + \beta \left| g^{(1)}(\tau) \right|^2 \quad , \qquad \text{(S10)}$$

where $\beta$ is a correction factor. For the diffusion of particles, $g^{(1)}(\tau)$ is related to the diffusion coefficient $D$ and the magnitude $q$ of the scattering vector

$$g^{(1)}(\tau) = e^{-Dq^2\tau} \quad . \qquad \text{(S11)}$$

$\beta$ is determined from fitting a stretched exponential function, according to $g^{(2)}(\tau) = \beta \cdot e^{-u \cdot \tau^v}$ to $g^{(2)}(\tau)$ for $8.8 \times 10^{-7} < \tau < 1.5 \times 10^{-5}$ s, where $u$ and $v$ are constants.

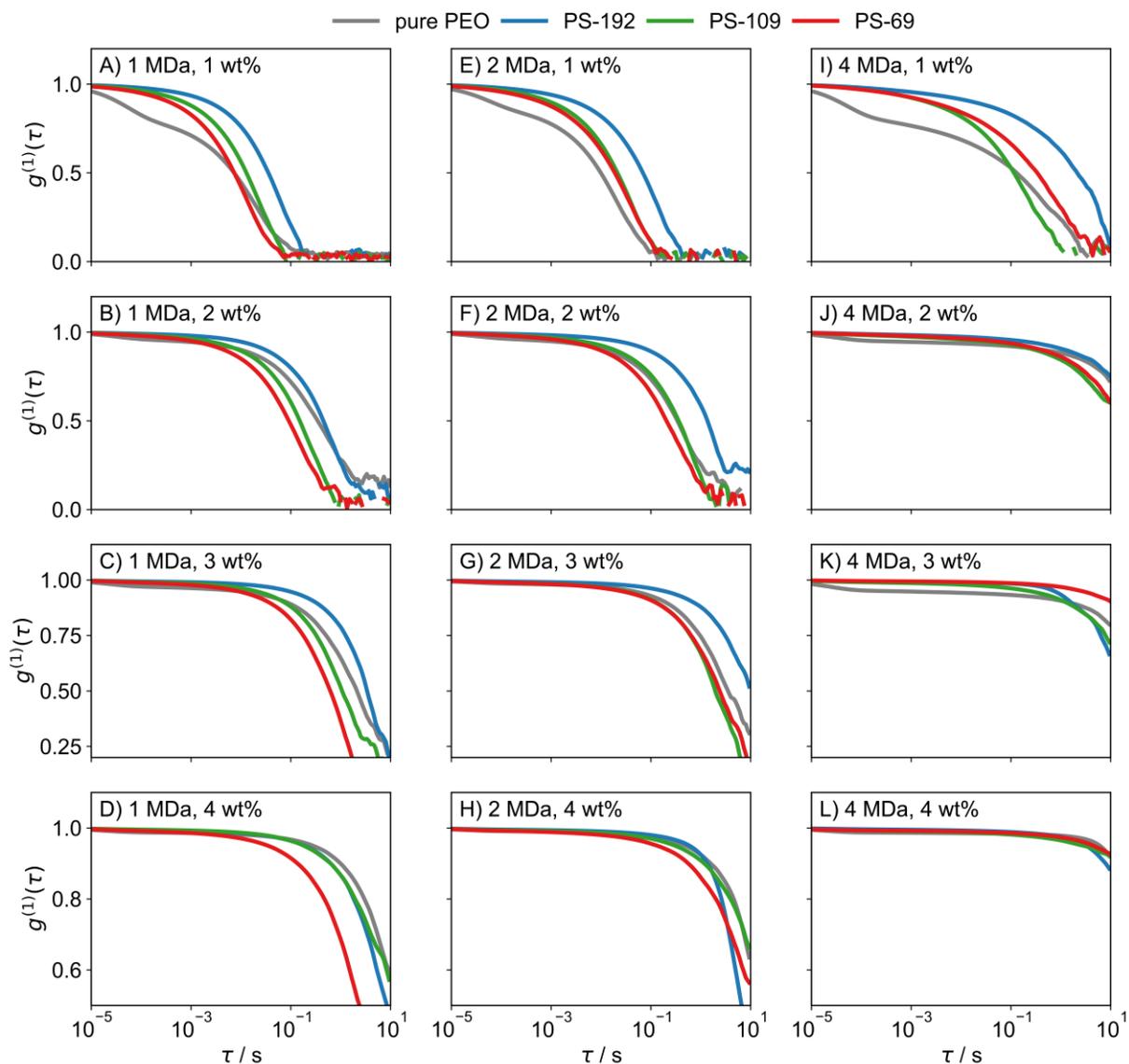

**Figure S4**. First-order correlation functions $g^{(1)}(\tau)$ for all PEO samples with three different tracer particle sizes as well as without particles.

The correlation functions $g^{(1)}(\tau)$ of all samples with the three different particle sizes as well as for the corresponding PEO samples without particles are shown in Figure S4. The slower dynamics, which occur at long lag times, are similar in both cases, indicating that the tracer particles follow the slow dynamics of the polymers. The short-time behavior looks significantly different with and without particles. In the absence of particles, faster dynamical processes taking place at smaller length scales are visible, which are masked as soon as particles are added. Using the mean-squared displacement (MSD) $\langle \Delta r^2(\tau) \rangle = 6D\tau$ and eqs S10 and S11, we obtain

$$\langle \Delta r^2(\tau) \rangle = -\frac{6}{q^2} \ln \sqrt{\frac{g^{(2)}(\tau)-1}{\beta}}, \qquad \text{(S12)}$$

which relates the experimentally accessible intensity auto-correlation function to the MSD of the tracer particles. We consider only lag times $\tau < 1$ s, for which $g^{(2)}(\tau) - 1 > 0.1$. Outside these bounds, the noise of $g^{(2)}(\tau)$ can lead to artefacts. Since the correlation function of particles trapped in low viscous samples decays faster, this means that the resulting MSDs will be cut off at a shorter lag time. MSDs calculated using eq S12 sometimes show $\langle \Delta r^2(0) \rangle \neq 0$, which stems from imperfect determination of $\beta$ by fitting. To remedy this, we perform a linear fit of the MSD for $10^{-6} < \tau < 5 \times 10^{-6}$ s and subtract the y-intercept from the data before further treatment of the data. The MSD is related to the frequency-dependent complex modulus $G^*(\omega)$ by the generalized Stokes-Einstein relation[24,25]

$$G^*(\omega) = \frac{k_B T}{\pi a i \omega \mathcal{F}_u\{\langle \Delta r^2(\tau) \rangle\}}, \qquad \text{(S13)}$$

where $i$ is the imaginary unit, $k_B$ is the Boltzmann constant, $T$ is the temperature, $a$ is the radius of the tracer particle, $\omega$ is the angular frequency, and $\mathcal{F}_u\{\langle \Delta r^2(\tau) \rangle\}$ denotes the single-sided Fourier transform of the MSD. Performing a numerical Fourier transform on the data is difficult due to the limited time range.[26] Instead, we adopted the procedure introduced by Mason *et al.*, where the MSD is expressed as a power law $\langle \Delta r^2(\tau) \rangle \approx \langle \Delta r^2(1/\omega) \rangle (\omega \tau)^{\alpha(\omega)}$, followed by an analytic Fourier transform.[27,28] The power law exponent $\alpha$ at time $\tau$ corresponds to the gradient of $\ln\langle \Delta r^2(\tau) \rangle$ with respect to $\ln \tau$

$$\alpha(\omega = 1/\tau) = \partial \ln\langle \Delta r^2(\tau) \rangle / \partial \ln \tau, \qquad \text{(S14)}$$

where we have used that the frequency is the inverse of the lag time. In viscoelastic fluids, $0 < \alpha < 1$ (0 corresponds to a purely elastic solid and 1 corresponds to a purely viscous liquid). Analytical Fourier transform of the local power law, together with eq S13, yields

$$G'(\omega) = |G^*(\omega)| \cos[\pi\alpha(\omega)/2] \,,$$

$$G''(\omega) = |G^*(\omega)| \sin[\pi\alpha(\omega)/2] \,, \qquad \text{(S15)}$$

where

$$|G^*(\omega)| = \frac{k_{\mathrm{B}}T}{\pi a \langle \Delta r^2(1/\omega) \rangle \Gamma[1+\alpha(\omega)]} \,. \qquad \text{(S16)}$$

Here, $\Gamma(z) = \int_0^\infty x^{z-1}\mathrm{e}^{-x}\mathrm{d}x$ denotes the Gamma function. More details about the derivation of eqs S15 and S16 and about the determination of $\alpha(\omega)$ are given in Supporting Information Section S6.

The Siegert relation, eq S10, is only valid for ergodic samples, for which the time-averaged scattering intensity is the same as the ensemble-averaged scattering intensity. This is no longer true if the particles are localized at fixed positions, as is the case in chemically crosslinked hydrogels. The PEO solutions used in this work should in principle be ergodic, since there are no permanent crosslinks. However, the measurement time might not be long enough. To verify the ergodicity of our PEO solutions, microrheology experiments were performed at ten different positions for very highly viscous samples, as shown in Figure S5.

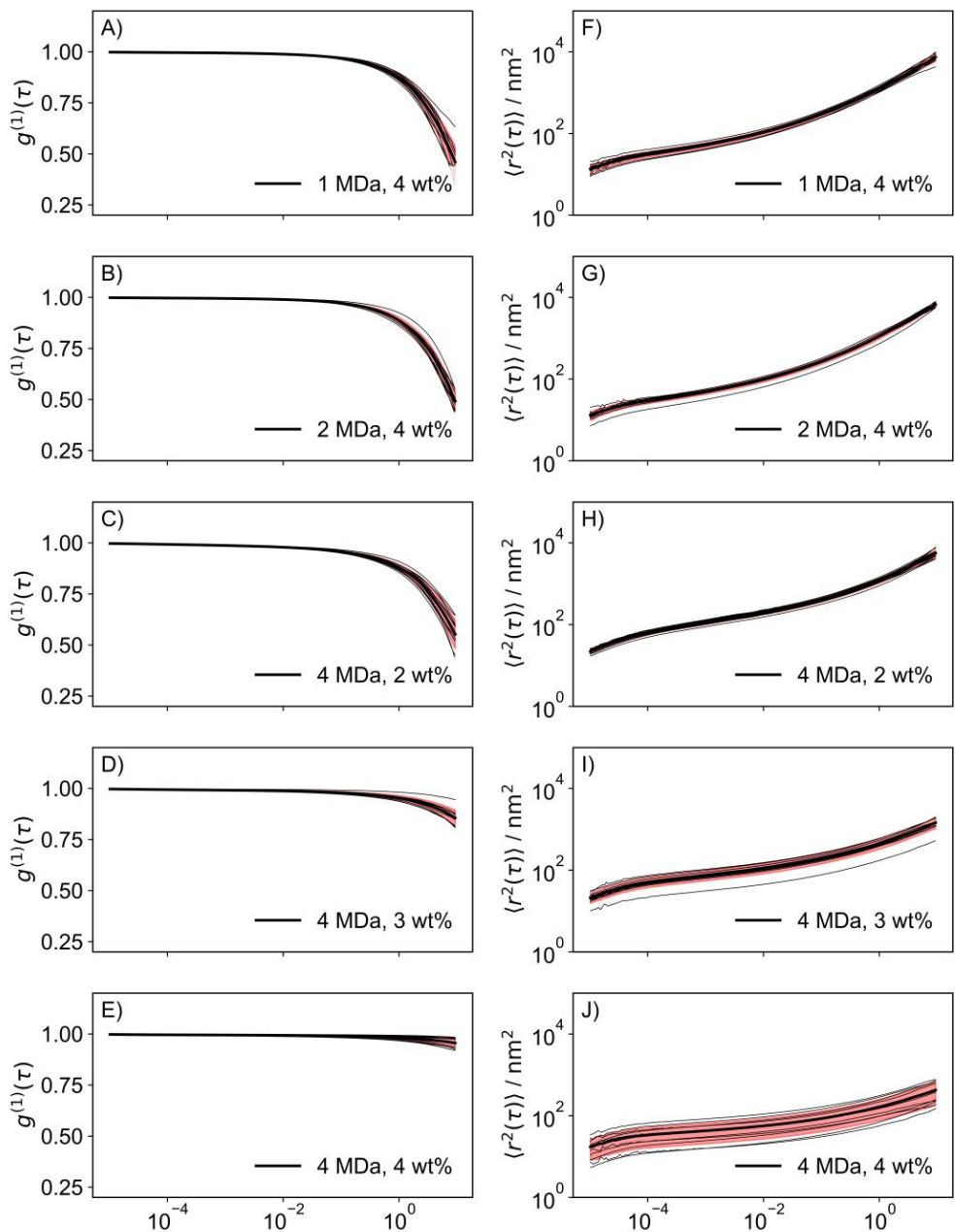

**Figure S5**. Microrheology experiments performed at ten different positions in the most highly viscous samples using PS-192 particles. In A)-E) the first order-correlation functions and in F)-J) the MSDs are shown. The fine black lines represent individual measurements, the thick black line indicates the average over all individual measurements, the shaded red area indicates the standard deviation.

For the 1 and 2 MDa samples, no significant variation of the correlation function with the measurement position was found, verifying that the ergodicity assumption holds. However, for the highest concentrated 4 MDa sample, signs of non-ergodic behavior were found. The non-ergodicity leads to an additional error for the two most highly viscous samples and explains why the error bars are so large for the 4 MDa samples in Figure 4 D and E in the main text. Since this is only a problem for the two most highly viscous samples, these effects have no influence on the main conclusions of our work.

## S6: Determination of the frequency-dependent power law exponent $\alpha(\omega)$

To determine $G^*(\omega)$ from eq S13, the MSD is expanded around $\tau = 1/\omega$, which yields $\langle \Delta r^2(\tau) \rangle \approx \langle \Delta r^2(1/\omega) \rangle (\omega \tau)^{\alpha(\omega)}$, where $\alpha$ is defined as the gradient $\alpha(\omega = 1/\tau) = \partial \ln\langle \Delta r^2(\tau) \rangle / \partial \ln \tau$. To find the gradient of a function $f$ that has at least 3 continuous derivatives for a non-homogeneous step-size, we seek to minimize the error $h_i$ between the true gradient and its estimate from a linear combination of neighbouring points. We minimize the consistency error $\epsilon_i$ between the true first derivative $f_i^{(1)} = \frac{df}{dx}\big|_{x=x_i}$ and its estimate from a linear combination of the neighboring data points with uneven spacings $h_s$ and $h_d$ [29–31]

$$\epsilon_i = f_i^{(1)} - [\alpha f(x_i) + \beta f(x_i + h_d) + \gamma f(x_i - h_s)]. \qquad (S17)$$

For the terms of the neighboring points, we substitute the first three terms of the Taylor expansions

$$f(x_i + h_d) = f(x_i) + h_d f^{(1)}(x_i) + \frac{h_d^2}{2} f^{(2)}(x_i) + \cdots \qquad , \qquad (S18)$$

$$f(x_i - h_s) = f(x_i) - h_s f^{(1)}(x_i) + \frac{h_s^2}{2} f^{(2)}(x_i) \mp \cdots \qquad , \qquad (S19)$$

to obtain

$$\epsilon_i = f_i^{(1)} - \left[ (\alpha + \beta + \gamma)f(x_i) + (\beta h_d - \gamma h_s)f^{(1)}(x_i) + \left( \frac{\beta h_d^2}{2} + \frac{\gamma h_s^2}{2} \right)f^{(2)}(x_i) \right]$$ .

(S20)

To estimate the first derivate $f_i^{(1)}$ we have to solve the linear system of equations

$$\alpha + \beta + \gamma = 0,$$

$$\beta h_d - \gamma h_s = 1,$$

$$\beta h_d^2 + \gamma h_s^2 = 0, \qquad \text{(S21)}$$

which yields for the approximation $\hat{f}_i^{(1)}$

$$\hat{f}_i^{(1)} \approx \frac{h_s^2 f(x_i + h_d) + (h_d^2 - h_s^2)f(x_i) - h_d^2(fx_i - h_s)}{h_s h_d (h_d + h_s)} + \mathcal{O}\left( \frac{h_d h_s^2 + h_s h_d^2}{h_d + h_s} \right).$$

(S22)

The results for the gradient $\alpha(\omega)$ are shown in Figure S6. Next, we evaluate the Fourier transform of the MSD algebraically from the expansion to find

$$i\omega \mathcal{F}_u \{\langle \Delta r^2(\tau) \rangle\} \approx \langle \Delta r^2(1/\omega) \rangle \Gamma[1 + \alpha(\omega)]i^{-\alpha(\omega)} \qquad \text{(S23)}$$

where $\Gamma$ is the gamma function. Substituting eq S23 into eq S13 and using Euler's equation, we obtain eq S15 and S16.

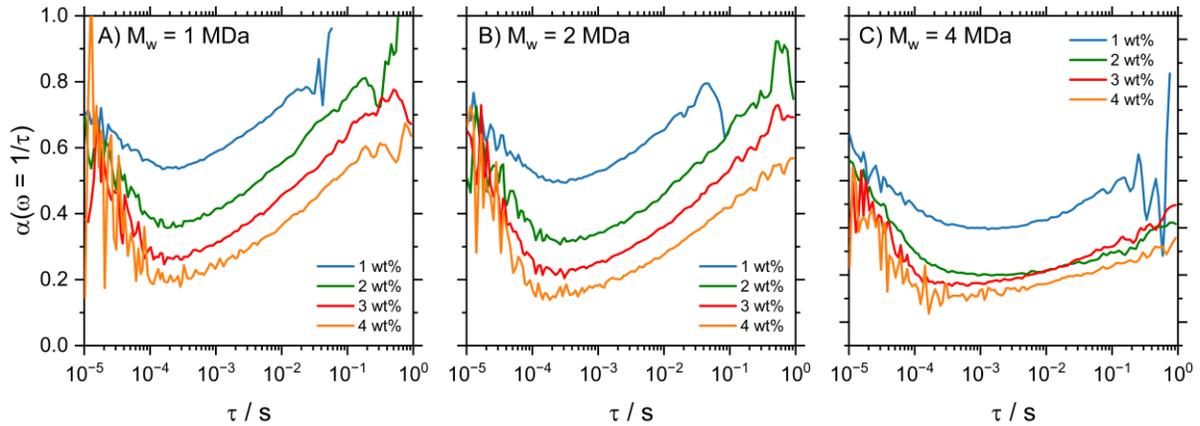

**Figure S6**. Logarithmic slope $\alpha(\omega)$ for the PS-109 samples.

**S7: Determination of $\eta_{\text{solv}}$ and $\eta_{\text{micro}}$ from microrheology**

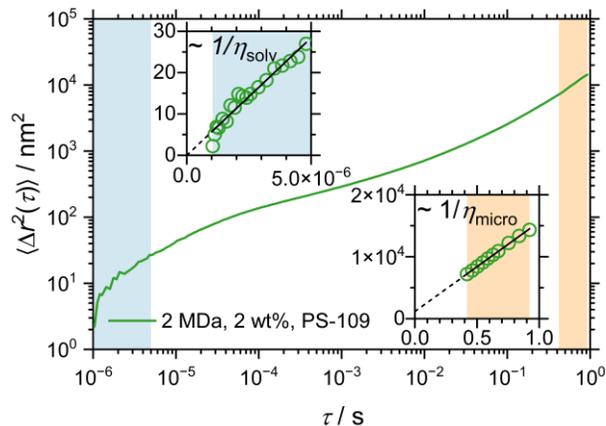

**Figure S7**. The MSD of tracer particles embedded in a viscoelastic fluid exhibits two linear diffusive regions at very short and very long times that are characterized by viscosities $\eta_{\text{solv}}$ and $\eta_{\text{micro}}$, respectively. The procedure is shown here for a 2 wt% solution of 2 MDa PEO from microrheology using PS-109 particles. The linear fits are shown by full lines in the insets and the broken lines extrapolate the fit to $\tau = 0$.

The mean-squared displacement (MSD) of tracer particles inside a viscoelastic medium can be divided into three separate regimes. At short lag times, the MSD is determined by the free diffusion of the particle through the solvent of viscosity $\eta_{\text{solv}}$ before its motion becomes influenced by the polymer matrix. At intermediate lag times, there is a subdiffusive plateau, which contains information about the frequency-dependent viscoelastic behavior of the sample. At very long lag times, the MSD becomes diffusive again, this time being governed by the zero-shear viscosity of the polymer matrix $\eta_{\text{micro}}$. $\eta_{\text{solv}}$ and $\eta_{\text{micro}}$ are determined from the slopes of the linear fits of the short- and long-time behavior, as shown in Figure S7, using $\langle \Delta r^2(\tau) \rangle = 6D\tau + b$ and $D = \frac{k_B T}{6\pi a \eta}$. For the linear fit of the short-time behavior, $b$ is assumed to vanish, $b = 0$. For the linear fit of the long-time behavior, we use the last 10 data points of each data set.

**S8: Determination of the shift factor**

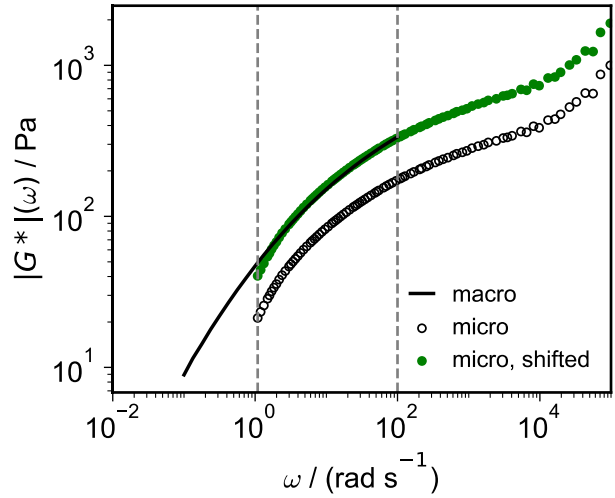

**Figure S8**. Example of the procedure to obtain the shift factor $\gamma_s$ between the micro- and macrorheological data $|G^*| = \sqrt{(G')^2 + (G'')^2}$ (for PS-192, $M_w = 2$ MDa and $c = 4$ wt%) in Figure 3B in the main text. The microrheology data are shifted such as to minimize the deviation between the data in the overlapping frequency range (denoted by grey vertical lines) $|G^*|_{macro} - |G^*_{micro,shifted}|$, where $|G^*_{micro,shifted}| = \gamma_s |G^*_{micro}|$.

The microrheological results of the dynamic moduli are shifted by a factor $\gamma_s$ to achieve agreement with the macrorheological data, i.e., $|G^*|_{\text{macro}} = |G^*_{\text{micro,shifted}}|$, where $|G^*_{\text{micro,shifted}}| = \gamma_s |G^*_{\text{micro}}|$ and $|G^*| = \sqrt{(G')^2 + (G'')^2}$. An example is given in Figure S8. Since for micro- and macrorheology data the frequency range of the data differs, first, the overlapping range is found (denoted by grey vertical lines in Figure S8). Then, the microrheological data in this range is interpolated by cubic splines. Ultimately, a least-square fit to minimize the distance between both datasets $|G^*|_{\text{macro}} - \gamma_s |G^*|_{\text{micro}}$ is done. The shift values for all concentrations, molecular weights, and tracer particles are summarized in Figure 3C in the main text. All adjusted microrheological data is given in Figure S9-S11.

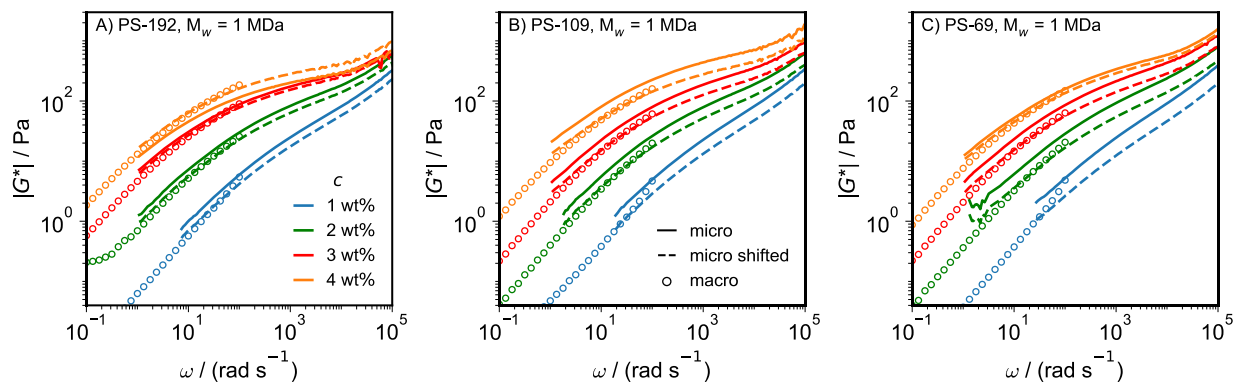

**Figure S9**. Comparison of the macrorheology data (circles) with the original (solid lines) and the shifted microrheology data (broken lines) for PEO solutions with $M_w$ = 1 MDa.

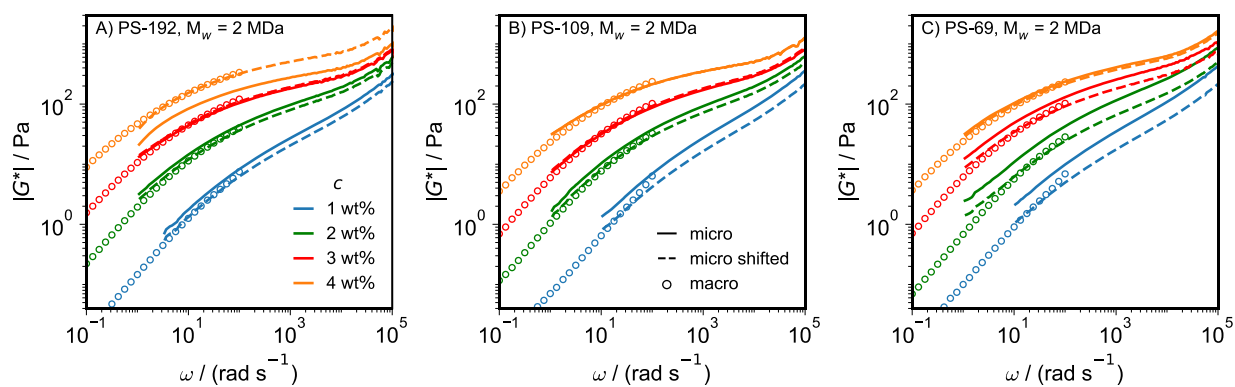

**Figure S10**. Comparison of the macrorheology data (circles) with the original (solid lines) and the shifted microrheology data (broken lines) for PEO solutions with $M_w$ = 2 MDa.

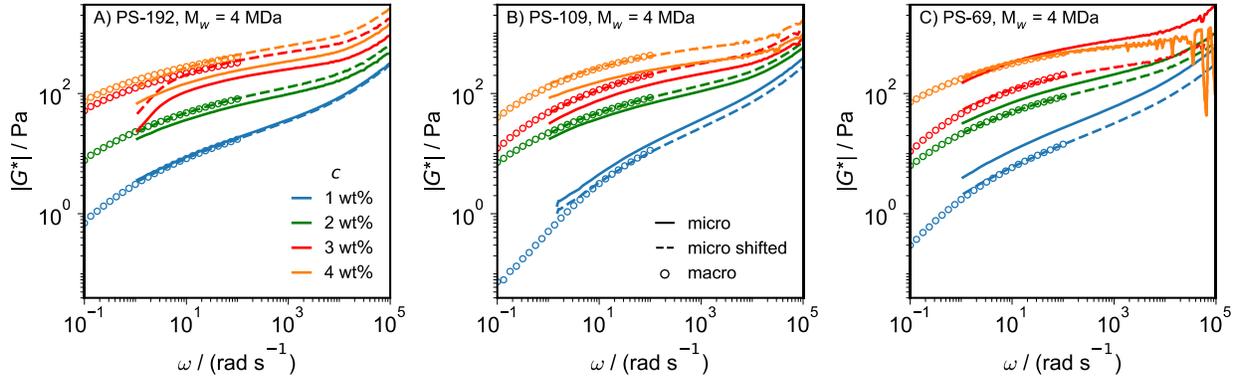

**Figure S11**. Comparison of the macrorheology data (circles) with the original (solid lines) and the shifted microrheology data (broken lines) for PEO solutions with $M_w$ = 4 MDa.

It becomes evident in Figure S8 and in the adjusted data in Figure S9-S11, that the slopes of both datasets do not always match perfectly in the overlapping range, especially for the low-concentration samples. Although we assume the same constant shift for $G'$ and $G''$, a significant improvement of the agreement between microrheology and macrorheology data for both $G'$ and $G''$ data is observed in all cases, as we demonstrate in Figure S12 for $M_w$ = 2 MDa, which justifies our model. This becomes even more visible in Figure S13, where we show the ratios between micro- and macrorheology, for $G'$ and $G''$ individually. An improvement can be seen in each case, but we observe that the ratios do not agree perfectly with 1 after shifting.

In Figure S14, the loss tangent tan($\delta$) = $G''/G'$ for all macro- and microrheology measurements is shown. In principle, tan($\delta$) of macro- and microrheology should always match, even if their corresponding $|G^*|$ are shifted against each other, because the correction factor $\gamma_s$, which acts equally on both $G'$ and $G''$, cancels out in tan($\delta$). The tan($\delta$) data demonstrate considerable deviations as the results from macrorheology experiments are always larger than the corresponding microrheology data, which cannot be remedied by the correction factor $\gamma_s$. Exceptions are the 3 and 4 wt%, 4 MDa, PS-192 samples, where tan($\delta$) from microrheology is larger than that of macrorheology. Here, most likely the microrheology results are erroneous due to the very slow relaxation of the DLS auto-correlation function of the large particles.

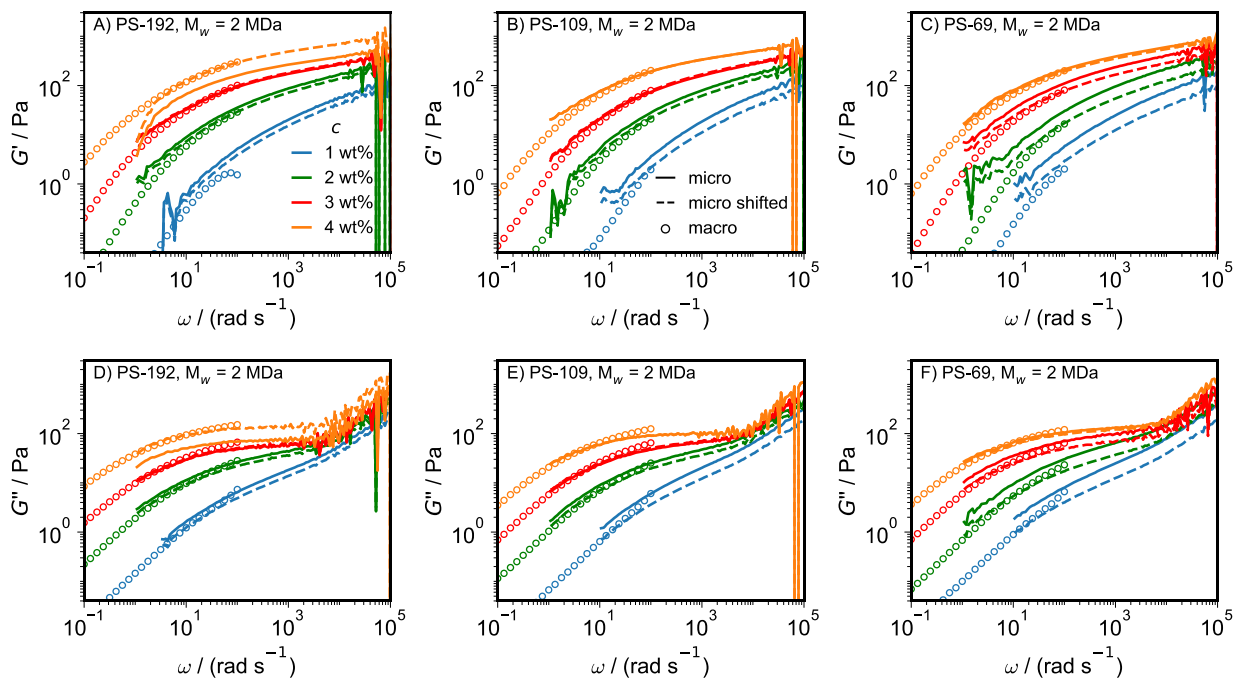

**Figure S12**. Viscoelastic moduli $G$' and $G$'' from macro- (circles) and microrheology (lines) experiments with $M_w$ = 2 MDa, together with the microrheology data that is shifted by the correction factor $\gamma_s$ (broken lines).

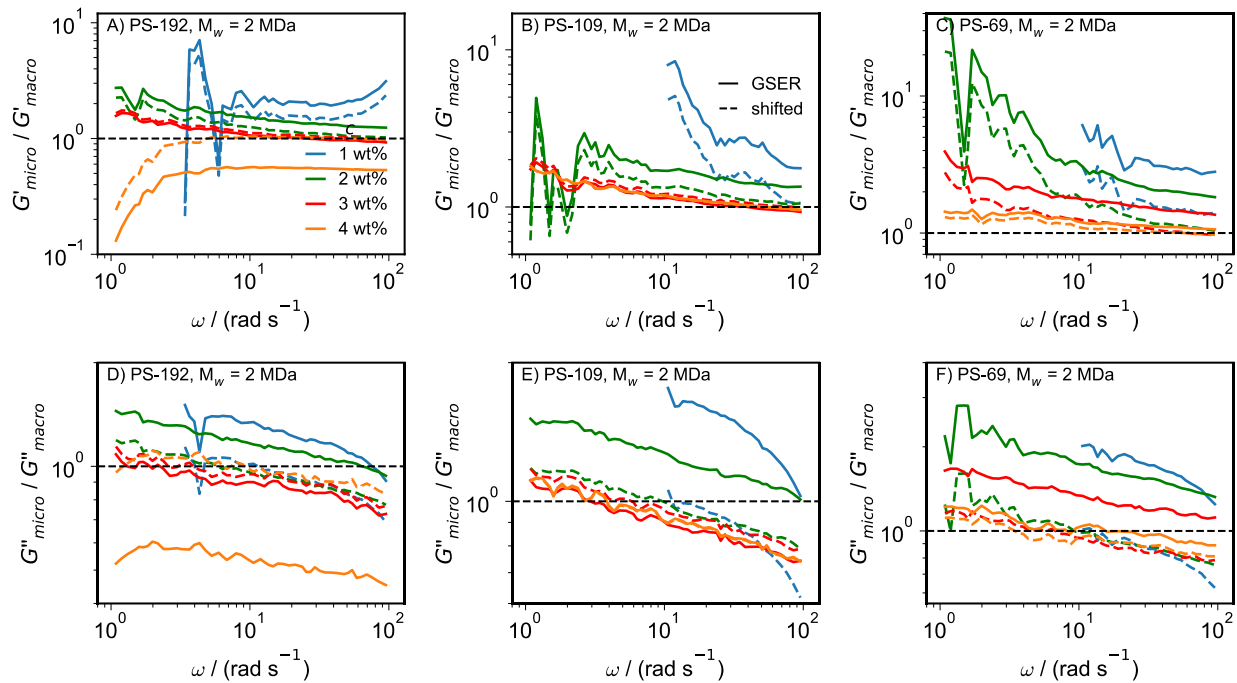

**Figure S13**. Ratios between the micro- and macrorheology experiments for $G$' and $G$'' with $M_w$ = 2 MDa (solid lines), compared with the ratio for the microrheology data that is shifted by the correction factor $\gamma_s$ (broken lines).

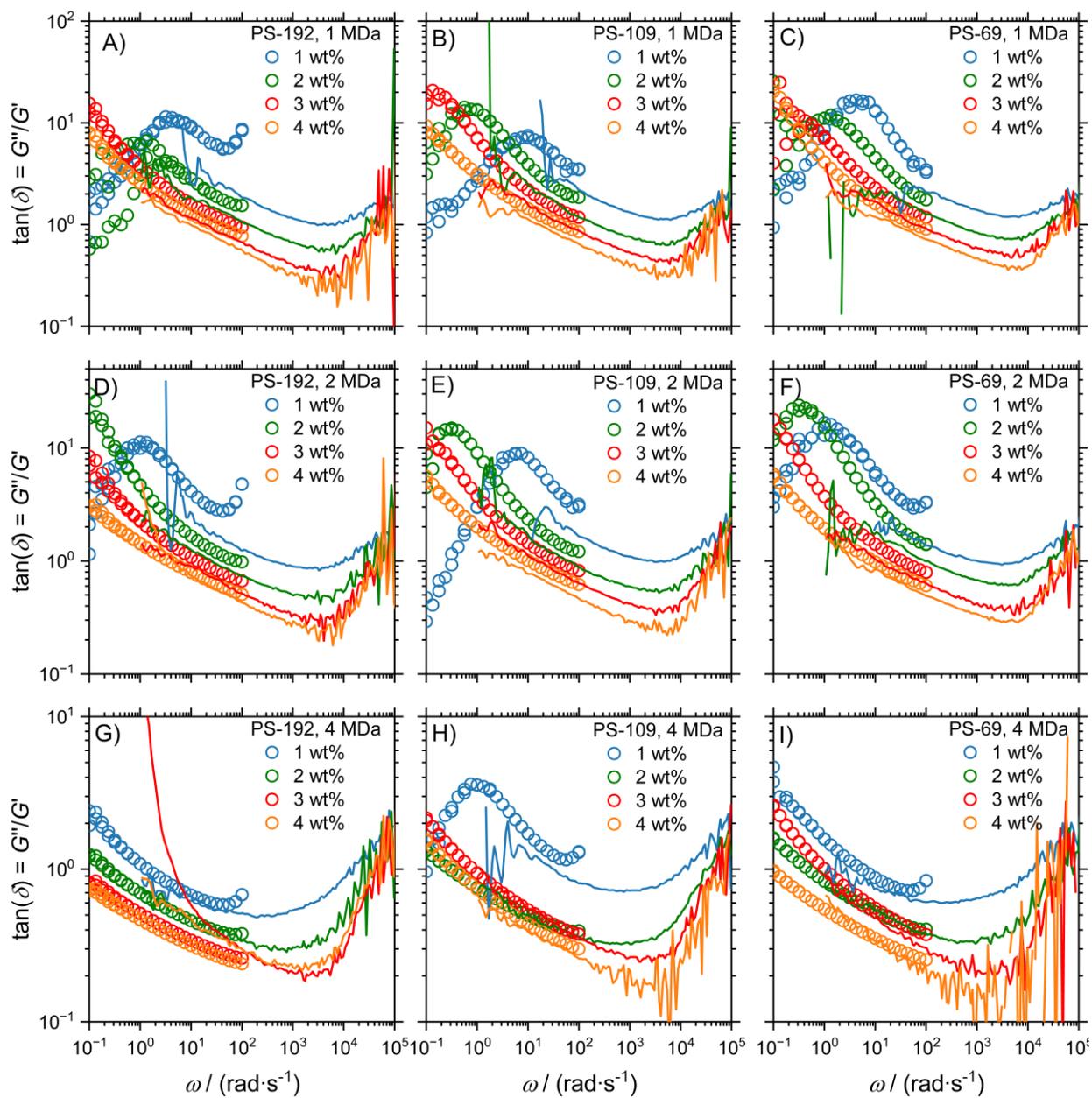

**Figure S14**. Ratio tan($\delta$) = $G''/G'$ of all macro- (circles) and microrheology (lines) experiments.

## S9: Estimation of molecular PEO radius

To estimate the molecular radius of a PEO chain, $R_{PEO}$, we approximate its shape as a cylinder of radius $R_{PEO}$. The cylinder length is given by the length of a PEO monomer unit, $a_0 = 0.356$ nm.[7] The mass contained inside that cylinder is given by $m_{cyl} = M_{mono}/N_A$, where $M_{mono}$ is the molar mass of the repeating unit and $N_A$ is Avogadro's number. The closest packing of cylinders in three dimensions is equal to the closest packing of circles in two dimensions and equal to a volume fraction of $\phi_{cyl} = 0.9069$. The density of a PEO melt $\rho_{melt}$ is around 1.13 g/cm$^3$. The density is given by

$$\rho_{melt} = \frac{m}{V} = \frac{m_{cyl}\phi_{cyl}}{V_{cyl}} = \frac{M_{mono}\phi_{cyl}}{N_A\pi R_{PEO}^2 a_0}. \qquad \text{(S24)}$$

Here, $V_{cyl}$ is the cylinder volume. By rearranging eq S24 and inserting the values for $a_0$ and the other constants, we find that $R_{PEO} = 0.229$ nm.

## S10: Model for effective solvent viscosity

In the main text, we explain the increased solvent viscosity in Figure 3A due to the formation of interfacial water layers with increased viscosity around the polymers. In Figure S15 we schematically illustrate the system along different directions. We envision the polymers as curved cylinders (shown as black solid lines or spheres), which are surrounded by interfacial water layers (broken lines). The spherical tracer (green sphere), which on short time scales freely diffuses through the solution, senses a combination of the bulk water viscosity $\eta_w$ and the interfacial viscosity $\eta_i$. The stress in the polymer solution, which results from the opposing shearing of two polymers (as illustrated in y- or z-direction), can be assumed to be constant in space and time in the stationary case and equals the friction force between individual water layers. In the semi-dilute regime, shearing of polymers with water in between cannot be assumed in a simplified way such as layered planar plate geometries.[32] Rather, we assume shearing to result from the average of parallel and perpendicular shearing of the bulk and interfacial water layers.

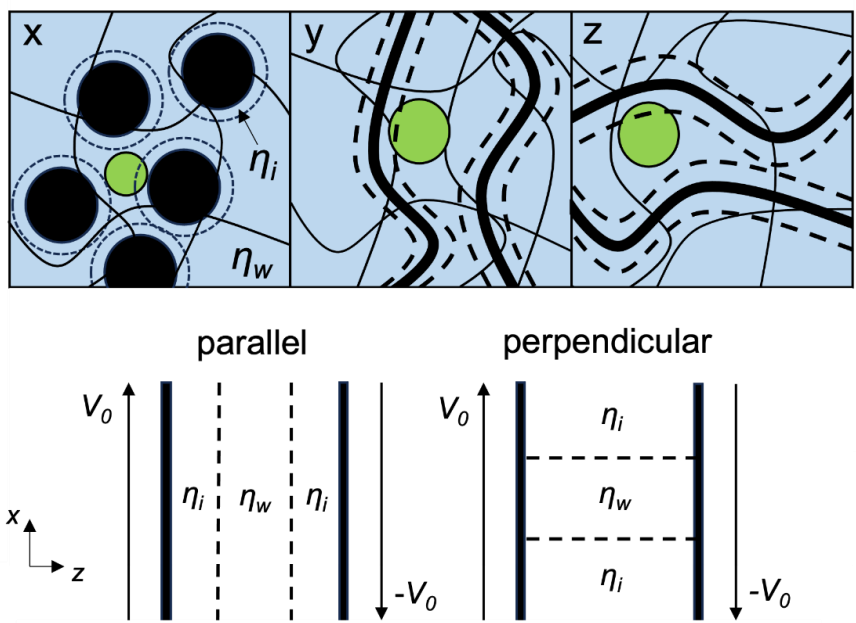

**Figure S15**. Model for the increased solvent viscosity used in Figure 3A in the main text. The upper scheme illustrates the semi-dilute solution including polymers (black spheres or solid lines) that are surrounded by interfacial layers (broken lines) with increased water viscosity $\eta_i$. We schematically show the diffusion of a spherical tracer particle (green sphere) through the solvent in different directions. Note that the sizes of the polymers and spheres in different directions are not drawn to scale. The system undergoes shearing of interfacial and bulk water layers in different directions, as we show in the lower scheme. The movement of polymers in opposite x-directions with velocity $v_0$ causes either a shearing of parallel or perpendicular interfacial and bulk water layers.

In the lower scheme of Figure S15, we illustrate the simplified model. The shearing of two parallel polymers in opposite x-direction with constant velocity $v_0$ causes a velocity along x that depends on the z position, $v_x(z)$, which is determined by the viscosity profile $\eta(z)$, the z-dependet friction force is given by $F_f \sim \eta(z)\,\partial_z v_x(z)$.[32] If the interfacial and bulk water layers are arranged in a parallel series, the solvent viscosity is the sum of the inverse of the individual layer viscosities, i.e., $1/\eta^{\parallel} = \phi_i/\eta_i + (1-\phi_i)/\eta_w$, where $\phi_i$ is the volume fraction of interfacial water. Due to the entanglement of polymers, the layers can also be arranged perpendicular to each other. Now the gradient of the velocity is constant, and the viscosity results from the sum, i.e., $\eta^{\perp} = \phi_i\eta_i + (1-\phi_i)\eta_w$. The effective solvent viscosity felt by a tracer particle is a sum of the viscosities from parallel and perpendicular sheared water layers. In the limit of low fractions, i.e., $\phi_i \to 0$, the viscosities behave as $\eta^{\parallel} = \eta_w + \phi_i(\eta_i - \eta_w)\eta_w/\eta_i$ and $\eta^{\perp} = \eta_w + \phi_i(\eta_i - \eta_w)$. Since we assume that $\eta_w \ll \eta_i$, it transpires that perpendicular components will dominate the average solvent viscosity felt by the tracer particle, which leads to eq 4 in the man text, i.e., $\eta_{\text{solv}} = \phi_i\eta_i + (1-\phi_i)\eta_w$, as a model for the effective solvent viscosity.

Using a value of $d$ = 0.4 nm for the thickness of the interfacial layer, the fit of eq 4 yields an interfacial viscosity of $\eta_i = (27.17 \pm 0.74)$ mPa s, which is in good agreement with the results from Netz *et al*.,[7] providing experimental verification of the simulation results and an explanation for the increased solvent viscosity determined using the tracer particles. When taking $d$ = 0.6 nm and 0.8 nm, we obtain the alternative estimates $\eta_i = (16.02 \pm 0.43)$ mPa s and $(10.71 \pm 0.28)$ mPa s, respectively. We can also determine the distance between the PEO cylinders by assuming a parallel arrangement. For concentrations of 1, 2, 3, and 4 wt%, we find distances of 4.26, 2.86, 2.24, and 1.86 nm after subtracting twice the radius of the polymer $2R_{\text{PEO}}$. These distances are of the same order as the distance between the two parallel surfaces used in the simulations in ref. [32].

## S11: Steady-shear experiments

The viscosity data from steady-shear experiments (see Figure S16) were fitted with the Cross model to determine the linear-response zero-shear viscosity $\eta_{\text{macro}}$. In the Cross model, the shear-rate-dependent non-linear viscosity $\eta_{\text{nl}}$ varies between the viscosity $\eta_{\text{macro}}$, obtained for $\dot{\gamma} \to 0$, and the infinite-shear viscosity $\eta_{\infty}$, obtained in the hypothetical limit $\dot{\gamma} \to \infty$, as

$$\eta_{\text{nl}} = \eta_{\infty} + \frac{\eta_{\text{macro}} - \eta_{\infty}}{1 + (k\dot{\gamma})^m}, \qquad \text{(S25)}$$

where $k$ is a characteristic crossover time and $m$ describes the sharpness or cooperativity of the shear-thinning transition. The fits are shown in Figure S16. For the very low viscous 1 wt% solutions, only shear rates higher than $1 \text{ s}^{-1}$ were considered. The infinite shear plateau characterized by $\eta_{\infty}$ is not reached in the considered shear rate range. Therefore $\eta_{\infty}$ was kept fixed at the viscosity of water at 25 °C, which is 0.89 mPa s.

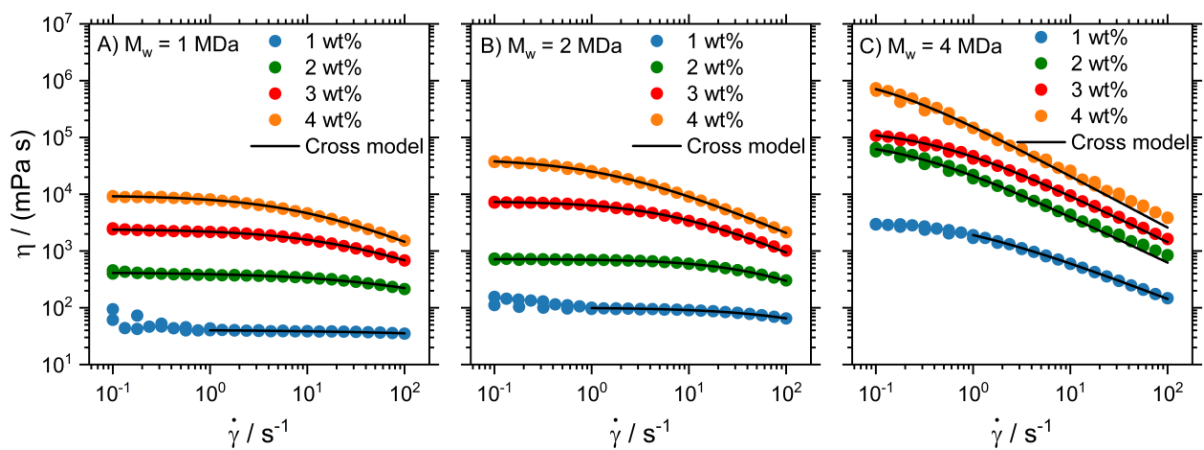

**Figure S16**. Viscosity data from steady-shear experiments for the PEO solutions. The data were fitted with the Cross model.

## S12: Transient and compressibility effects in PEO solutions

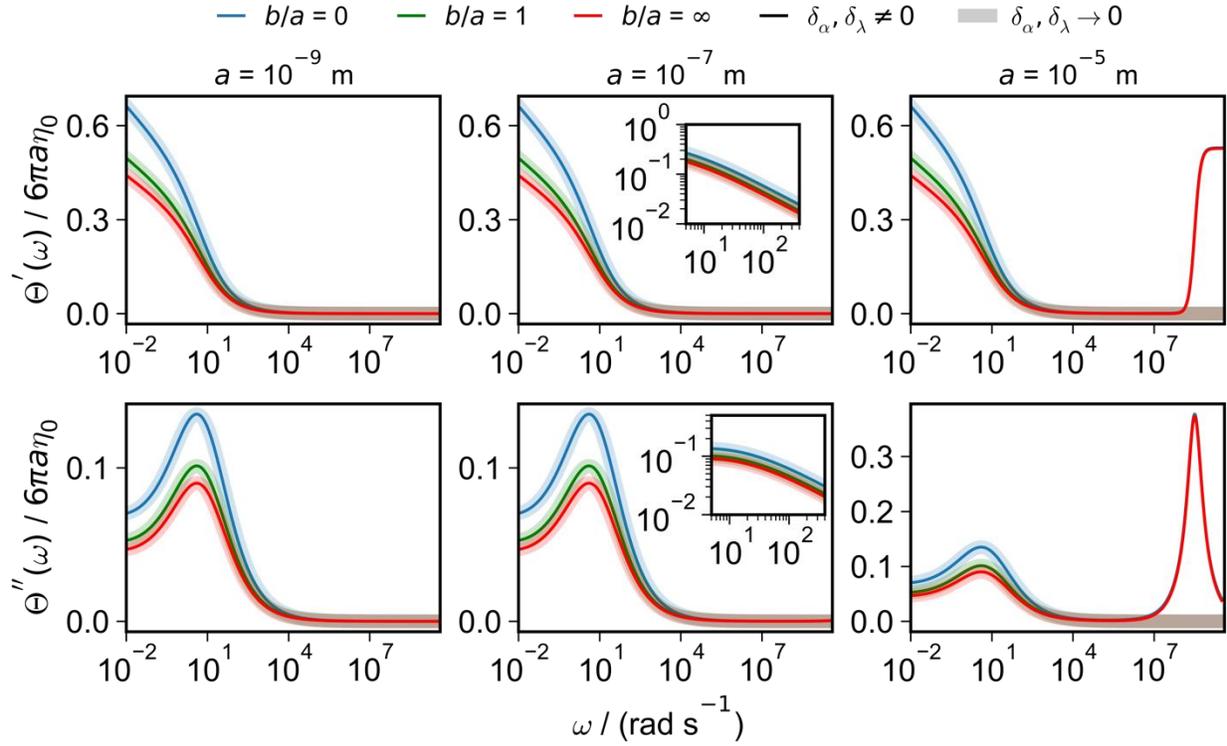

**Figure S17.** Frequency-dependent friction coefficient $\Theta(\omega) = \Theta'(\omega) + i\Theta''(\omega)$ of a sphere calculated using eq S26 for different radii $a$ and slip lengths $b$. For the shear viscoelasticity $\eta_{ve}(\omega) = G^*(\omega)/i\omega$, we used the FMM-fit for the 1 MDa and 3 wt% PEO solution (see Supporting Information Section S4 for details), and for the volume viscoelasticity we used $\zeta_{ve}(\omega) = \eta_{ve}(\omega)$. Additionally a speed of sound of $c = 2250$ m/s [33] and density $\rho_0 = 1.13 \cdot 10^3$ kg/$m^3$ is chosen. Shadows denote the steady incompressible limit, corresponding to $\delta_\alpha \to 0$ and $c \to \infty$ ($\delta_\lambda \to 0$).

The GSER in eq 2 in the main text is derived from the Navier-Stokes equation using the Stokes' approximation, neglecting transient and compressibility effects in the momentum and continuity equation. [34,35] These equations can be solved to obtain the velocity and pressure field of the fluid.

The total frictional force $F_\Theta$ of a moving sphere with radius $a$ in the fluid (using a stick boundary condition) with velocity amplitude $v$ is found to fulfill the Stokes law, i.e., $F_\Theta = \Theta v$, where the friction coefficient is given by $\Theta(\omega) = 6\pi a \eta_{ve}(\omega)$, which leads, by assuming that the friction coefficient and the viscoelasticity $\eta_{ve}$ are both frequency-dependent,[25] to the GSER for the dynamic moduli in eq S13 using the relation between the friction coefficient and the mean-squared displacement,[24,25] i.e., $\Theta(\omega) = \frac{6k_B T}{(i\omega)^2 \mathcal{F}_u\{\langle \Delta r^2(\tau)\rangle\}}$ , and $G^*(\omega) = i\omega \eta_{ve}(\omega)$. A much more comprehensive solution of the transient Stokes equation for compressible fluids and for slip boundary conditions at the spherical surface has been carried out by Erbaş $et$ $al.$.[36] The expression of the frequency-dependent friction coefficient includes correction terms due to transient and compression effects and is given by

$$\Theta(\omega) = \frac{4\pi \eta_{ve}(\omega) a W^{-1}}{3}\{(1+\delta_\lambda)(9 + 9\delta_\alpha + \delta_\alpha^2)(1 + 2\,\hat{b}) + (1 + \delta_\alpha)[2\delta_\lambda^2(1 + 2\,\hat{b}) +$$

$$\hat{b}\delta_\alpha^2(1+\delta_\lambda)]\}, \qquad (S26)$$

where $W$ is defined as

$$W = (2 + 2\delta_\lambda + \delta_\lambda^2)[1 + \hat{b}(3 + \delta_\alpha)] + (1 + \delta_\alpha)(1 + 2\,\hat{b})\delta_\lambda^2/\delta_\alpha^2. \qquad (S27)$$

The dimensionless slip length $\hat{b} = b/a$ describes finite slip at the spherical surface. The dimensionless decay constants $\delta_\alpha$ and $\delta_\lambda$ describe the propagation of shear and compression waves in the fluid and are defined by

$$\delta_\alpha^2 = -\frac{i\omega a^2 \rho_0}{\eta_{ve}(\omega)}, \qquad (S28)$$

and

$$\delta_\lambda^2 = \frac{-i\omega a^2 \rho_0}{\frac{4\eta_{ve}(\omega)}{3} + \zeta_{ve}(\omega) + i\rho_0 c^2/\omega}, \qquad (S29)$$

where $\rho_0$ is the mean fluid density, $c$ the speed of sound, and $\eta_{ve}(\omega)$ and $\zeta_{ve}(\omega)$ the frequency-dependent shear and volume viscoelasticities. In the limit $b \to 0$, $\delta_\alpha \to 0$ and $c \to \infty$, the friction coefficient converges to the Stokes law expression, i.e., $\Theta(\omega) = 6\pi a \eta_{ve}(\omega)$, and the standard GSER in eq S13 is recovered.

In Figure S17, we show the frequency-dependent friction coefficient $\Theta(\omega) = \Theta'(\omega) + i\Theta''(\omega)$ of a sphere calculated using eq S26 for different radii $a$ and slip lengths $b$. For the shear viscoelasticity $\eta_{ve}(\omega) = G^*(\omega)/i\omega$ the FMM-fit for the 1 MDa and 3 wt% PEO solution (see Supporting Information Section S4 for details) is used, and for the volume viscosity we choose $\zeta_{ve}(\omega) = \eta_{ve}(\omega)$. Additionally, the speed of sound is $c = 2250$ m/s [33] and the density is $\rho_0 = 1.13 \cdot 10^3$ kg/m$^3$. Shadows denote the steady incompressible limit, i.e., $\delta_\alpha \to 0$ and $c \to \infty$ ($\delta_\lambda \to 0$). A difference between shadows and solid lines would therefore correspond to a deviation of the standard GSER from the full solution. In the experimental frequency range of $10^{-1} < \omega < 10^5$ rad/s no deviations are observable for all sphere sizes, which clearly shows that transient and compression effects are unimportant for PEO solutions in the experimentally relevant frequency range. In Figure S18 we compare the friction for different radii and negligible slip, i.e., $b = 0$. As visible in Figure S17 and Figure S18, deviations from the GSER become evident for radii exceeding 10 $\mu$m, which lies above the investigated tracer size. For high frequencies, a plateau in the real part and a peak in the imaginary part are visible. These effects correspond to the propagation of compression waves in the medium, where the resonance peak in the imaginary part equals the inverse time the wave needs to travel over a distance corresponding to the tracer size.[36] These effects move into the experimentally relevant frequency range for radii around 1 mm (Figure S18), which is clearly above the tracer sizes used in our experimental setup.

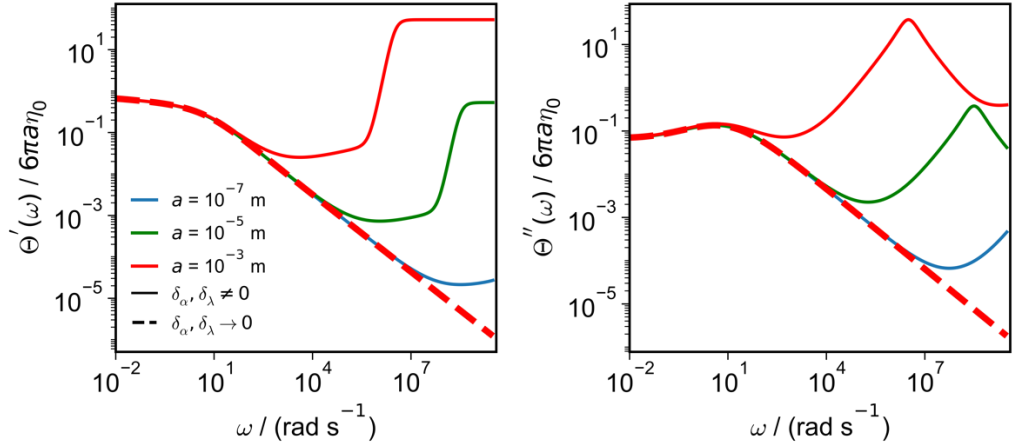

**Figure S18**. Frequency-dependent friction coefficient $\Theta(\omega) = \Theta'(\omega) + i\Theta''(\omega)$ of a moving sphere calculated using eq S26 for different radii $a$ and using $b = 0$. Broken lines denote the steady incompressible limit, i.e., $\delta_\alpha \to 0$ and $c \to \infty$ ($\delta_\lambda \to 0$).

As seen in Figure S17, the amplitude of the friction coefficient depends on the slip parameter $b$. For $\delta_\alpha \to 0$ and $c \to \infty$, the friction coefficient in eq S26 becomes

$$\Theta(\omega) = 6\pi\eta_{\text{ve}}(\omega)a\frac{1+2\hat{b}}{1+3\hat{b}}, \quad \text{(S30)}$$

which results in a modified effective shear viscoelasticity $\eta_{\text{ve}}^{\text{eff}}(\omega) = \eta_{\text{ve}}(\omega)\frac{1+2\hat{b}}{1+3\hat{b}}$. Clearly, for $b > 0$, the dynamic moduli $G^*(\omega) = i\omega\eta_{\text{ve}}^{\text{eff}}(\omega)$ are decreased compared to the standard GSER, which does not explain shifts between macro- and microrheology smaller than 1. This finding forms the motivation to use the shell model instead.

## S13: Small-angle neutron scattering of PEO solutions

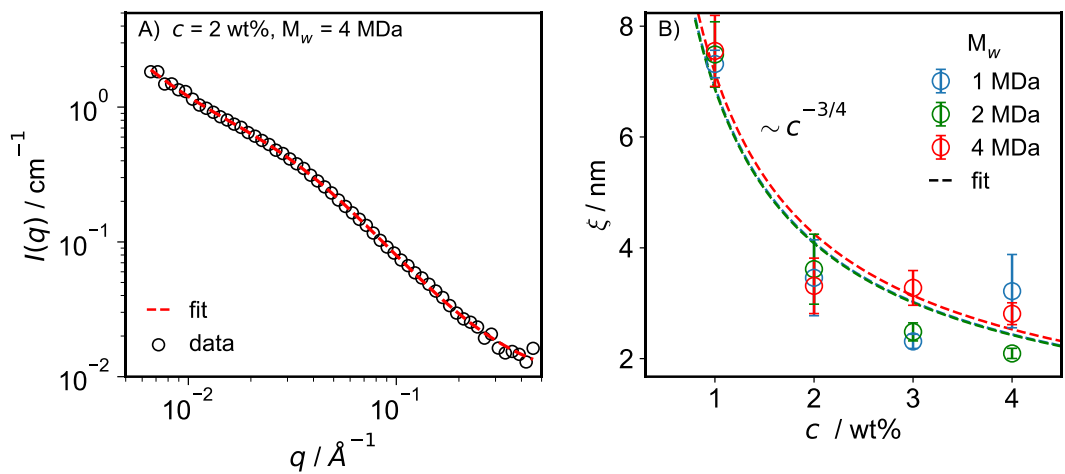

**Figure S19.** A) Exemplary scattering intensity $I(q)$ obtained from SANS of PEO solutions in $D_2O$ together with a fit according to the Hammouda-model in eq S31. B) Estimated mesh size of the PEO solutions from the model in eq S31 as a function of the polymer concentration $c$ and for different molecular weights. We additionally show fits according to eq S32, which illustrate that the mesh size decreases with increasing concentration according to scaling theory.

Measuring the scattering intensity $I(q)$ in dependence on the scattering vector $q$ with small-angle neutron scattering (SANS) enables insights into the hydrogel structure down to the nanoscale. SANS experiments were performed on the LARMOR instrument at ISIS Pulsed Neutron and Muon Source (Didcot, United Kingdom) using a temperature-controlled sample changer and rectangular quartz cuvettes of 2 mm thickness. For details on sample preparation, we refer to Supporting Information Section S2. The temperature was fixed at 25°C. Neutron wavelengths of 0.9 to 13 Å were used simultaneously (time-of-flight), yielding a total q-range of $4.5 \times 10^{-3}$ to $6.7 \times 10^{-1}$ Å$^{-1}$. Data reduction was done using the MANTID software.[37] The raw intensity data were corrected for background scattering and weighted by the transmission of the sample.[38] The absolute scaling is performed using a secondary calibrated polymer blend sample.[39] Finally, the 2D data were radially averaged.

In Figure S19A we show an exemplary scattering profile $I(q)$ of a PEO-solution in $D_2O$. Structural features such as heterogeneities or distributions of mesh sizes can be investigated by fitting the data with an empirical correlation-length model, suggested by Hammouda *et al.*.[40] The scattering intensity is given by

$$I(q) = \frac{I_a}{q^{k_p}} + \frac{I_b}{1+(q\xi)^{k_l}} + I_c , \quad \text{(S31)}$$

where $k_p$ and $k_l$ are the Porod and Lorentzian exponent, respectively. $I_c$ is the background scattering. The correlation length (or mesh size) $\xi$ can be obtained by fitting the data with eq S31, where we demonstrate a good agreement between data and model in Figure S19A. In Figure S19B we show the fitted mesh size in dependence on the sample's polymer concentration and molecular weight. The mesh size overall ranges between values 2 – 8 nm and decreases with increasing concentration, which is expected according to scaling theory,[6,41] i.e.,

$$\xi = R_g \left(\frac{c}{c^*}\right)^{-\frac{3}{4}} . \quad \text{(S32)}$$

Fixing the radius of gyration to the empirical relation, i.e., $R_g/\text{Å} = 0.215(M_w/(\text{g mol}^{-1}))^{0.583\pm0.031}$, given in Table S1, we can estimate the overlap concentration by fitting the data in Figure S19B with eq S32. We observe a good agreement (see broken lines in Figure S19B) and obtain the values $c^* = 0.048 \pm 0.001$ wt%, $0.028 \pm 0.002$ wt% and $0.017 \pm 0.002$ wt% for 1 MDa, 2 MDa and 4 MDa, respectively. The other constants of the fit in eq S31 are given in Table S2. For some samples, the fit errors for parameters $I_a$ and $k_p$, found in the first term in eq S31 (associated with heterogeneities) are rather large. This does, however, not affect the determination of the correlation length $\xi$, whose error is small in all cases.

**Table S2.** Results from the Hammouda-model fit in eq S32 to the SANS data.

| 1 MDa | 1 wt% | 2 wt% | 3 wt% | 4 wt% |
|---|---|---|---|---|
| $I_a$ / cm$^{-1}$ | $(2.84 \pm 0.28) \cdot 10^{-13}$ | $(1.77 \pm 6.27) \cdot 10^{-4}$ | $(5.44 \pm 4.62) \cdot 10^{-5}$ | $(2.09 \pm 5.06) \cdot 10^{-4}$ |
| $I_b$ / cm$^{-1}$ | $1.36 \pm 0.05$ | $0.7 \pm 0.32$ | $0.63 \pm 0.06$ | $0.60 \pm 0.24$ |
| $I_c$ / cm$^{-1}$ | $(4.12 \pm 0.04) \cdot 10^{-3}$ | $(7.8 \pm 0.01) \cdot 10^{-3}$ | $(1.42 \pm 0.77) \cdot 10^{-2}$ | $(7.44 \pm 1.12) \cdot 10^{-3}$ |
| $k_p$ | $5.27 \pm 2.08$ | $1.72 \pm 0.66$ | $2.0 \pm 0.17$ | $1.7 \pm 0.46$ |
| $k_l$ | $1.81 \pm 0.03$ | $1.83 \pm 0.15$ | $1.87 \pm 0.05$ | $1.68 \pm 0.11$ |
| $\xi$ / Å | $73.17 \pm 0.25$ | $34.59 \pm 0.69$ | $23.11 \pm 0.13$ | $32.18 \pm 0.67$ |

| 2 MDa | 1 wt% | 2 wt% | 3 wt% | 4 wt% |
|---|---|---|---|---|
| $I_a$ / cm$^{-1}$ | $(9.97 \pm 2.34) \cdot 10^{-13}$ | $(1.79 \pm 5.79) \cdot 10^{-4}$ | $(4.13 \pm 3.86) \cdot 10^{-5}$ | $(7.56 \pm 4.19) \cdot 10^{-6}$ |
| $I_b$ / cm$^{-1}$ | $1.39 \pm 0.13$ | $0.72 \pm 0.29$ | $0.66 \pm 0.07$ | $0.66 \pm 0.04$ |
| $I_c$ / cm$^{-1}$ | $(4.36 \pm 0.36) \cdot 10^{-3}$ | $(6.14 \pm 0.67) \cdot 10^{-3}$ | $(1.35 \pm 0.01) \cdot 10^{-2}$ | $(1.25 \pm 0.01) \cdot 10^{-2}$ |
| $k_p$ | $5.08 \pm 4.63$ | $1.7 \pm 0.6$ | $2.08 \pm 0.19$ | $2.47 \pm 0.11$ |
| $k_l$ | $1.801 \pm 0.04$ | $1.8 \pm 0.13$ | $1.88 \pm 0.06$ | $1.80 \pm 0.04$ |
| $\xi$ / Å | $74.89 \pm 0.59$ | $36.15 \pm 0.64$ | $24.83 \pm 0.17$ | $20.94 \pm 0.10$ |

| 4 MDa | 1 wt% | 2 wt% | 3 wt% | 4 wt% |
|---|---|---|---|---|
| $I_a$ / cm$^{-1}$ | $(3.98 \pm 2.82) \cdot 10^{-5}$ | $(2.19 \pm 4.44) \cdot 10^{-4}$ | $(1.71 \pm 2.79) \cdot 10^{-5}$ | $(1.12 \pm 0.98) \cdot 10^{-5}$ |
| $I_b$ / cm$^{-1}$ | $1.33 \pm 3.37$ | $0.64 \pm 0.23$ | $0.78 \pm 0.12$ | $0.78 \pm 0.07$ |
| $I_c$ / cm$^{-1}$ | $(3.15 \pm 1.13) \cdot 10^{-3}$ | $(8.99 \pm 0.62) \cdot 10^{-3}$ | $(1.02 \pm 0.01) \cdot 10^{-2}$ | $(9.85 \pm 0.09) \cdot 10^{-3}$ |
| $k_p$ | $1.7 \pm 12.06$ | $1.73 \pm 0.38$ | $2.23 \pm 0.32$ | $2.37 \pm 0.17$ |
| $k_l$ | $1.77 \pm 0.67$ | $1.91 \pm 0.14$ | $1.86 \pm 0.06$ | $1.74 \pm 0.05$ |
| $\xi$ / Å | $75.53 \pm 6.41$ | $33.13 \pm 0.51$ | $32.77 \pm 0.32$ | $28.10 \pm 0.20$ |

## S14: Mesh size of a cubic polymer network

We assume that the polymers form a cubic lattice where the chains with contour length $L_0 = a_0 N$ are stretched along the edges, and the mesh size $\xi_{\text{cubic}}$ is the distance between two nodes in the lattice. In the lattice, the monomeric number density of polymers $\phi_{\text{m}}$ is given by the ratio between the number of monomers per cube and the cubic volume, i.e., $\phi_{\text{m}} = 3(\frac{\xi_{\text{cubic}}}{a_0})/\xi_{\text{cubic}}^3 = 3/(a_0 \xi_{\text{cubic}}^2)$. Consequently, the mesh size follows as $\xi_{\text{cubic}} = \left(\frac{3}{a_0 \phi_{\text{m}}}\right)^{1/2}$.

## S15: Derivation of the shell-model GSER

We start from the steady momentum and continuity equations for an incompressible fluid, which, for a homogenous fluid, are given by

$$\nabla \mathrm{p} = \eta \nabla^2 \boldsymbol{v} \,, \quad \text{(S33)}$$

$$\nabla \cdot \boldsymbol{v} = 0 \,, \quad \text{(S34)}$$

where $\boldsymbol{v}$ is the vectorial velocity field of the fluid, p is the pressure field, and $\eta$ is the shear viscosity. We use the two-layer model introduced by Fan *et al.*,[42] where a sphere of radius $a$ is surrounded by a shell of thickness $\Delta$ and local viscosity $\eta_{\text{shell}}$ (see Figure 4A in the main text for a schematic drawing), so the viscosity profile is given by

$$\eta(r) = \left\{ \begin{array}{ll} \eta_{\text{shell}}, & \text{for } a \leq R \leq a + \Delta, \\ \eta_{\text{macro}}, & \text{for } R > a + \Delta. \end{array} \right. \quad \text{(S35)}$$

Note that, contrary to the considerations of Fan *et al.*,[42] the local viscosity in our model can be greater than the bulk viscosity $\eta_{\text{macro}}$. The composite fluid dynamics of the system is described by separate equations for the inner and outer layer, with velocities $\boldsymbol{v}^i$ and $\boldsymbol{v}^o$, respectively, and are given by

$$\nabla \mathrm{p}^{(i)} = \nabla^2 \boldsymbol{v}^{(i)} \text{ and } \nabla \cdot \boldsymbol{v}^{(i)} = 0 \quad \text{for } 1 \leq r \leq 1 + \delta, \text{ (S36)}$$

$$\nabla \mathrm{p}^{(o)} = \kappa \nabla^2 \boldsymbol{v}^{(o)} \text{ and } \nabla \cdot \boldsymbol{v}^{(o)} = 0 \quad \text{for } r > 1 + \delta, \text{ (S37)}$$

where $\kappa = \eta_{\mathrm{shell}}/\eta_{\mathrm{macro}}$ and $\delta = \Delta/a$. Note that all equations are normalized such that velocities have the units of $U$ (velocity amplitude of the fluid, here in the z-direction), i.e., $\boldsymbol{v} \sim U$, positions the units of the spherical radius, i.e., $\boldsymbol{r} \sim a$, and viscosities the units of the bulk value $\eta_{\mathrm{macro}}$, such that $p \sim \eta_{\mathrm{shell}} U/a$, and so that the momentum eqs S36 and S37 are both divided by $\eta_{\mathrm{shell}} U/a$. We assume that $\kappa$ and $\Delta$ are constant in time and space.

The eqs S36 and S37 are solved by finding their Stokes stream functions $\psi^{(i,o)}$ in spherical coordinates[43]

$$\boldsymbol{v}^{(i,o)} = v_{\mathrm{r}}^{(i,o)} \hat{\boldsymbol{e}}_r + v_{\phi}^{(i,o)} \hat{\boldsymbol{e}}_\phi = -\frac{1}{r^2 sin\phi} \frac{\partial \psi^{(i,o)}}{\partial \phi} \hat{\boldsymbol{e}}_r + \frac{1}{r \, sin\phi} \frac{\partial \psi^{(i,o)}}{\partial \mathrm{r}} \hat{\boldsymbol{e}}_\phi \ , \quad \text{(S38)}$$

which is done by using the Ansatz $\Omega \, \psi^{(i,o)} = 0$, where $\Omega$ is given by

$$\Omega = \left[ \frac{\partial^2}{\partial r^2} + \frac{sin\phi}{r^2} \frac{\partial}{\partial \phi} \left( \frac{1}{sin\,\phi} \frac{\partial}{\partial \phi} \right) \right]^2 . \quad \text{(S39)}$$

Note that we follow the solution scheme of Fan *et. al.*,[42] but with the extension of non-restricted values for $\kappa$. A solution for the inner and outer stream function is $\psi^{(i,o)} = sin^2 \phi \ f^{(i,o)}(r)$ where for $f^{(i,o)}(r)$ we have

$$\frac{8f^{(i,o)}}{r^4} = -\frac{8f^{(i,o)\prime}}{r^3} + \frac{4f^{(i,o)\prime\prime}}{r^2} - f^{(i,o)\prime\prime\prime\prime}, \quad \text{(S40)}$$

which has the solution

$$f^{(i,o)}(r) = \mathrm{A}^{(i,o)} \mathrm{r}^4 + \mathrm{B}^{(i,o)} r^2 + \mathrm{C}^{(i,o)} r + \frac{\mathrm{D}^{(i,o)}}{r} \ . \quad \text{(S41)}$$

Thus, we have 8 unknowns in the complete form of the stream functions, i.e., $\psi^{(i,o)} = sin^2 \phi \left( A^{(i,o)} r^4 + B^{(i,o)} r^2 + C^{(i,o)} r + \frac{D^{(i,o)}}{r} \right)$. The unknowns are calculated by using boundary conditions, which couple the inner and outer layers at $r = 1$ and $r = 1 + \delta$. The condition

$$\text{v}_\text{r}^{(i,o)} \to 0 \quad \text{as} \ \ r \to \infty, \quad \text{(S42)}$$

dictates $A^{(o)} = 0$ and $B^{(o)} = 0$. Assuming continuity, the shear stresses $\sigma$ and the total stresses are equal at the shell boundary

$$\sigma_\text{r,\phi}^{(i)} = \sigma_\text{r,\phi}^{(o)} \quad \text{at} \ \ r = 1 + \delta, \quad \text{(S43)}$$

$$-\text{p}^{(i)} + \sigma_\text{r,r}^{(i)} = -\text{p}^{(o)} + \sigma_\text{r,r}^{(o)} \quad \text{at} \ \ r = 1 + \delta, \quad \text{(S44)}$$

and the velocity fields as well

$$\text{v}_\text{r}^{(i)} = \text{v}_\text{r}^{(o)} \quad \text{at} \ \ r = 1 + \delta, \quad \text{(S45)}$$

$$\text{v}_\phi^{(i)} = \text{v}_\phi^{(o)} \quad \text{at} \ \ r = 1 + \delta. \quad \text{(S46)}$$

Additionally, we assume negligible slip

$$\text{v}_\text{r}^{(i)} = \cos\phi \quad \text{at} \ \ r = 1, \quad \text{(S47)}$$

$$\text{v}_\phi^{(i)} = -\sin\phi \quad \text{at} \ \ r = 1. \quad \text{(S48)}$$

Eq S43 is written in spherical coordinates

$$\kappa\left[\text{r}\frac{\partial}{\partial \text{r}}\left(\frac{\text{v}_\phi^{(i)}}{\text{r}}\right) + \frac{\partial \text{v}_\text{r}^{(i)}}{\text{r}\,\partial\phi}\right] = \text{r}\frac{\partial}{\partial \text{r}}\left(\frac{\text{v}_\phi^{(o)}}{\text{r}}\right) + \frac{\partial \text{v}_\text{r}^{(o)}}{\text{r}\,\partial\phi}, \quad \text{(S49)}$$

which leads, by using the definition of the stream functions, to the relation

$$\kappa(1 + \delta)^5 A^{(i)} + \kappa D^{(i)} = D^{(o)}, \quad \text{(S50)}$$

Applying the stream functions in the integrated versions of the eqs S36 and S37 for the pressures $p^{(i,o)}$ with $p^{(i,o)}(\to \infty) = 0$ leads to

$$p^{(i)} = -20(1 + \delta)\cos\phi\, A^{(i)} - \frac{2\cos\phi}{(1+\delta)^2}C^{(i)} \quad \text{at} \ \ r = 1 + \delta, \quad \text{(S51)}$$

$$p^{(o)} = -\frac{2\cos\phi}{\kappa(1+\delta)^2}C^{(o)} \quad \text{at} \ \ r = 1 + \delta. \quad \text{(S52)}$$

Inserting eqs S51 and S52 into eq S44, we obtain

$$2(1+\delta)^5 A^{(i)} + (1+\delta)^2 C^{(i)} + 2D^{(i)} - \frac{(1+\delta)^2}{\kappa} C^{(o)} = \frac{2}{\kappa} D^{(o)}. \quad \text{(S53)}$$

Finally, connecting the conditions in eqs S45 – S48, it follows that

$$(1+\delta)^5 A^{(i)} + (1+\delta)^3 B^{(i)} + (1+\delta)^2 C^{(i)} + D^{(i)} - (1+\delta)^2 C^{(o)} = D^{(o)}, \quad \text{(S54)}$$

$$4(1+\delta)^5 A^{(i)} + 2(1+\delta)^3 B^{(i)} + (1+\delta)^2 C^{(i)} - D^{(i)} - (1+\delta)^2 C^{(o)} = D^{(o)}, \quad \text{(S55)}$$

$$A^{(i)} + B^{(i)} + C^{(i)} + D^{(i)} = -\frac{1}{2}, \quad \text{(S56)}$$

$$4A^{(i)} + 2B^{(i)} + C^{(i)} - D^{(i)} = -1. \quad \text{(S57)}$$

All unknown constants $A^{(i)}, B^{(i)}, C^{(i)}, D^{(i)}, C^{(o)}$ and $D^{(0)}$ are found by solving the system of eqs S50, S53 and S54-S57

$$A^{(i)} = 3(\kappa-1)(1+\delta)[-1+(1+\delta)^2]/2\Pi, \quad \text{(S58)}$$

$$B^{(i)} = -1[5(\delta_s+1)(\kappa-1) - 4(-1+\kappa)^2 - 3(\delta+1)^5(-3+\kappa+2\kappa^2)]/2\Pi, \quad \text{(S59)}$$

$$C^{(i)} = -3[(1+\delta)(2(\kappa-1)+(1+\delta)^5(2+3\kappa)]/2\Pi, \quad \text{(S60)}$$

$$D^{(i)} = -2[(1+\delta)^3(\kappa-1)+(\delta+1)^5(2+3\kappa))]/2\Pi, \quad \text{(S61)}$$

$$C^{(o)} = 3\kappa[2(\delta+1)(\kappa-1)+(\delta+1)^6(2+3\kappa)]/2\Pi, \quad \text{(S62)}$$

$$D^{(0)} = (1+\delta)^3\kappa[5(\delta+1)^3+2(\kappa-1)+3(\delta+1)^5(\kappa-1)]/2\Pi, \quad \text{(S63)}$$

where $\Pi$ is given by

$$\Pi = 9(\delta+1)(\kappa-1) - 10(\delta+1)^3(\kappa-1) + 4(\kappa-1)^2 + (\delta+1)^6(6\kappa+4) + 3(\delta+1)^5(\kappa+2\kappa^2-3). \quad \text{(S64)}$$

Using the solutions for the stream functions $\psi^{(i,o)}$, and accordingly for $v_{r,\phi}^{(i,o)}$ and $p_{r,\phi}^{(i,o)}$, we can compute the total drag force $F_\Theta$ by integrating the sum of shear stress and normal stress over the entire spherical surface.[42] Using eq S46, we arrive at

$$F_\Theta = 6\pi\eta_{\text{shell}} a U g_s \hat{e}_z, \quad \text{(S65)}$$

where $g_s$ is the correction factor given by

$$g_s = \frac{1}{\Pi}\left[2(2+3\kappa)\left(1+\frac{\Delta}{a}\right)^6 - 4(1-\kappa)\left(1+\frac{\Delta}{a}\right)\right]. \quad (S66)$$

Eq S65 is the Stokes law with friction coefficient $\Theta = 6\pi\eta_{\text{shell}}ag_s$. Note that we have assumed constant shear viscosities $\eta_{\text{shell}}$ and $\eta_{\text{macro}}$ in the derivation. To derive the Stokes relation between the frequency-dependent friction coefficient $\Theta(\omega)$ and the viscoelasticity $\eta_{\text{shell}}^{\text{ve}}(\omega)$, we would have to solve the transient Stokes as shown by Erbaş *et al.*,[36] here for the shell model. However, since we found in Supporting Information Section S12 that finite sphere radius and compressibility effects are negligible in our systems, we continue directly with the generalized Stokes law,[24,25] i.e., $\Theta(\omega) = 6\pi\eta_{\text{shell}}^{\text{ve}}(\omega)ag_s$.

Using the relation between friction coefficient and mean-squared displacement, i.e., $\Theta(\omega) = \frac{6k_BT}{(i\omega)^2\mathcal{F}_u\{\langle\Delta r^2(\tau)\rangle\}}$, we arrive at the shell-model GSER in eq 5 in the main text using that $\gamma_s^{-1} = \kappa g_s$, $\eta_{\text{shell}}^{\text{ve}}(\omega) = \kappa\eta_{\text{macro}}^{\text{ve}}(\omega)$, and $G_{\text{macro}}^*(\omega) = i\omega\eta_{\text{macro}}^{\text{ve}}(\omega)$. Consequently, comparing eq 2 and 6 in the main text, we obtain the relation between the macrorheology and the microrheology modulus as $|G^*|_{\text{macro}}(\omega) = \gamma_s(\Delta,\kappa)\cdot|G^*|_{\text{micro}}(\omega)$.

## S16: Effective tracer radii from experimental shifts

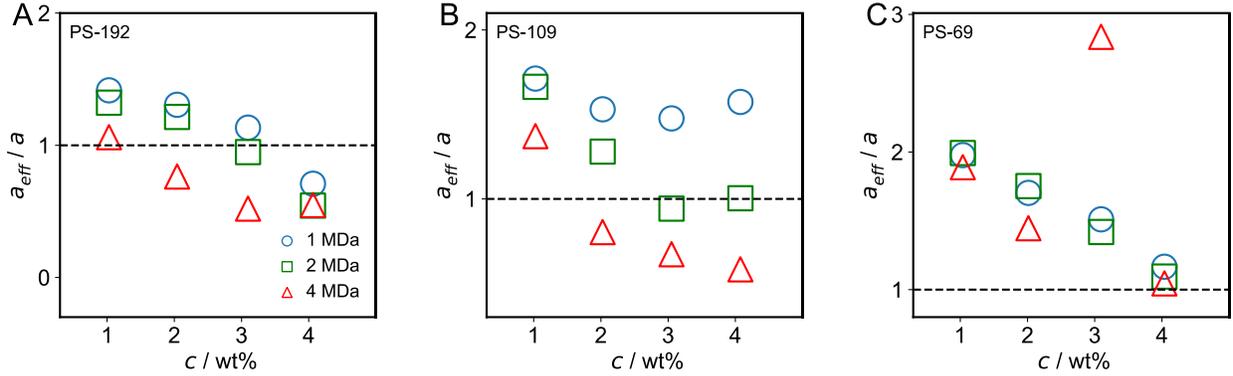

**Figure S20**. Effective radii $a_{\text{eff}}/a = 1/\gamma_s$, computed from the shift factors shown in Figure 3C in the main text, for different tracer-particle sizes and PEO molecular weights as a function of PEO concentration.

Deviations between macro- and microrheology data is quantified by a frequency-independent shift factor $\gamma_s$ according to $\left| G^*_{\text{micro,shifted}} \right| = \gamma_s |G^*_{\text{micro}}|$, which can be attributed to a slowdown or acceleration of the tracer particle dynamics in the hydrogel due to the formation of an interfacial layer around the particles. A relatively simple approach to explain these effects consists of modifying the effective hydrodynamic radius of a particle according to $a_{\text{eff}} = \epsilon_s a$, from which the friction coefficient follows as $\Theta_{\text{eff}}(\omega) = 6\pi\epsilon_s a \eta_{\text{ve}}(\omega)$ and the GSER as [44]

$$|G^*(\omega)| = \frac{k_{\text{B}}T}{\pi\epsilon_s a \langle \Delta r^2 (1/\omega)\rangle \Gamma[1+\alpha(\omega)]}. \quad (S67)$$

Consequently, the shift factor $\gamma_s$ according to $\left| G^*_{\text{micro,shifted}} \right| = \gamma_s |G^*_{\text{micro}}|$ is the inverse of the relative effective radius, i.e., $a_{\text{eff}}/a = 1/\gamma_s$. In Figure S20, we summarize the effective radii inferred from the shifts shown in Figure 3C in the main text. The values exhibit the same behavior as $\gamma_s$ but demonstrate that increased or reduced particle sizes up to twice or half the tracer size are required to explain the shifts.

**S17: Adjusting the dynamic moduli data using the shell model**

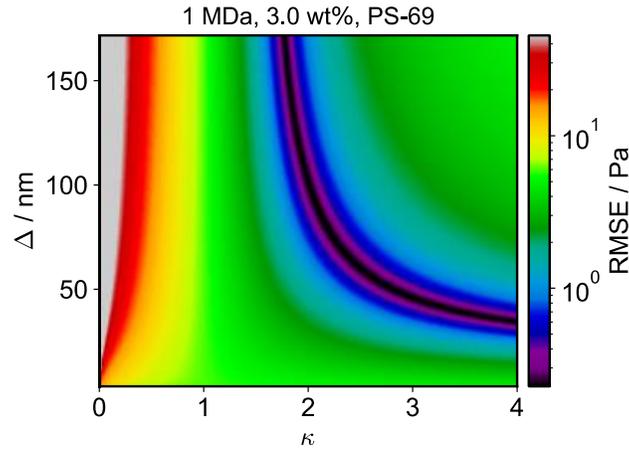

**Figure S21**. Root mean square error (RMSE) defined in eq S68 between $|G^*|_{macro}$ and $\gamma_s(\Delta, \kappa) \cdot |G^*|_{micro}$ for different combinations of $\Delta$ and $\kappa$. Here we show an exemplary result for the dataset (1 MDa, 3 wt%, PS-69).

In principle, we could try to determine the values for $\Delta$ and $\kappa$ for the shell model introduced in Supporting Information Section S15 by the same fitting procedure as proposed in Supporting Information Section S8, that means by minimizing the distance between the macrorheology and the corrected microrheology data, i.e., $|G^*|_{macro} - \gamma_s(\Delta, \kappa)|G^*|_{micro}$ and by using eqs 5 (main text), S64 and S66. However, finding meaningful fitting values using a least-squares fit is difficult, since a unique minimum for a certain combination of $\Delta$ and $\kappa$ does not exist, as we demonstrate in Figure S21. We define the RMSE between $|G^*|_{macro}$ and $\gamma_s(\Delta, \kappa)|G^*|_{micro}$ by

$$\text{RMSE} = \sqrt{\frac{1}{N}\sum_{i=0}^{N}(|G^*|_{\text{macro}}(\omega_i) - \gamma_s(\Delta,\kappa)|G^*|_{\text{micro}}(\omega_i))^2}, \quad (S68)$$

for $N$ overlapping data points of the macro- and microrheological datasets (1 MDa, 3 wt%, PS-69) at the frequencies $\omega_i$ and show the RMSE for different combinations of $\Delta$ and $\kappa$. A curved path through the parameter space at which the RMSE is minimal is observed. Thus, fitting both parameters simultaneously is impossible. For this reason, we keep one of the two parameters fixed, here the shell thickness, and identify the minimizing $\Delta$-value for all datasets on average. In Figure S22, we display the error for optimal $\kappa$ between $|G^*|_{\text{macro}}$ and $\gamma_s(\Delta, \kappa) \cdot |G^*|_{\text{micro}}$ averaged over datasets with different concentrations and molecular weights (A), over different concentrations and tracer radii (B), and over different molecular weights and tracer radii (C). We see that saturation of the RMSE values occur above certain layer thicknesses $\Delta_{min}$ (denoted by broken vertical lines) but at different locations depending on the radius, molecular weight, or concentration, respectively. When we plot the RMSE averaged over different polymer concentrations and polymer molecular weights as a function of the ratio of $\Delta$ and $R_e{}^{\text{ideal}}$, we see in (D) that the RMSE becomes constant above a universal layer thickness $\Delta_{min} \approx \frac{3}{5} R_e{}^{\text{ideal}}$, indicated by a vertical line, independent of the tracer particles radius, which is an empirical proof of our proposed scaling of $\Delta$ with $R_e{}^{\text{ideal}}$. In Figure S8-S11 in Supporting Information Section S8, we show a comparison of the macrorheology data with the original microrheology and the shifted microrheology data.

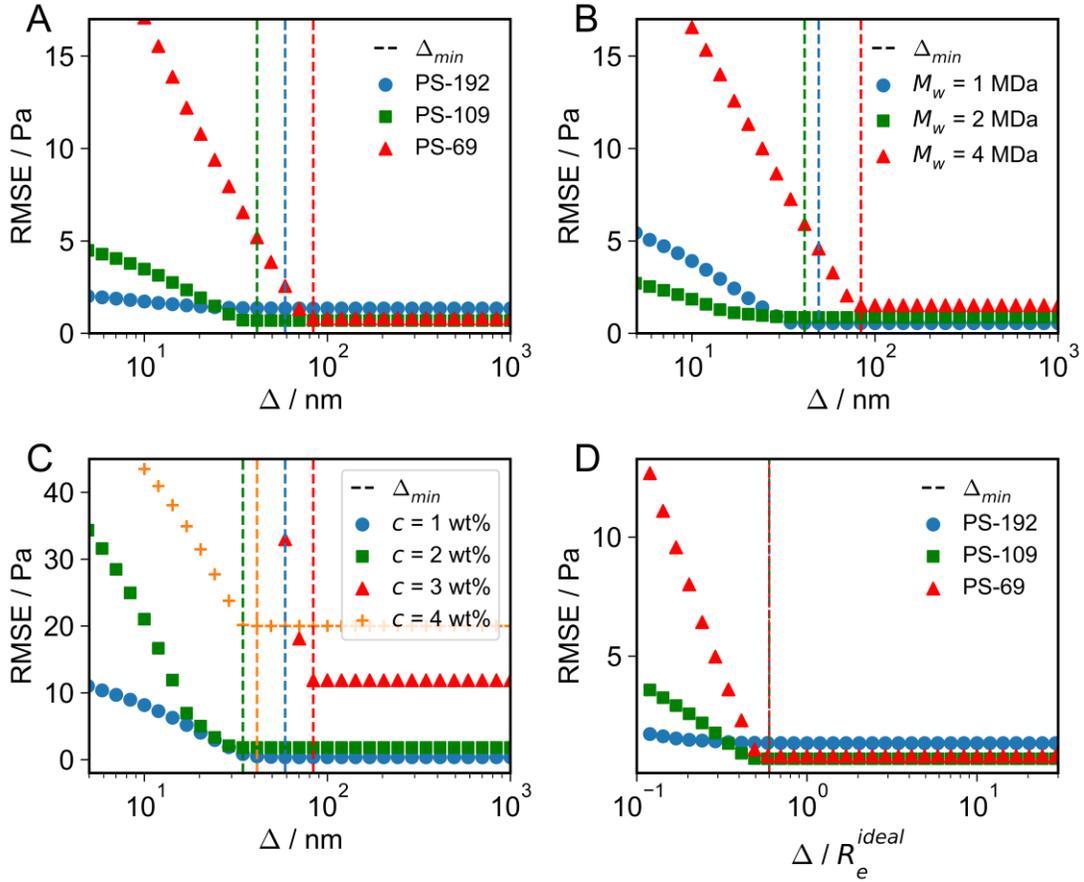

**Figure S22**. RMSE, as defined in eq S68, for fitting $\kappa$ in the relation $|G^*|_{macro} = \gamma_s(\Delta, \kappa)|G^*|_{micro}$, where we used eq 5 in the main text for the fit (same procedure as in Supporting Information Section S8), but with fixed shell thickness $\Delta$. The RMSE is averaged over datasets with different concentrations and molecular weights (A), over different concentrations and tracer radii (B), and over different molecular weights and tracer radii (C). (D) RMSE averaged over datasets with different concentrations and molecular weights and shown for different tracer radii, where $\Delta$ is chosen in terms of the end-to-end distance $R_e^{ideal}$, given in Table S1. Broken lines denote the convergence values of the fitting error, which is in (D) around $\Delta_{min} \approx \frac{3}{5} R_e^{ideal}$ for all radii, representing the value chosen to describe the experimentally determined shift factor.

## S18: Dependence of shell modulus on tracer size

We investigate whether the shell viscosity determined by rheology with different tracer particles follows the same power laws as the averaged trend shown in Figure 4C-E in the main text. First we display in Figure S23 the values $\kappa = G^*_{\text{shell}}(\omega)/G^*_{\text{macro}}(\omega)$ as a function of polymer concentration, which are already shown in Figure 4B in the main text. The values for PS-192 and PS-109 are shifted downwards compared to PS-69 and for high concentrations become smaller than 1, which presumably is due to polymer depletion effects around the tracer particles. In fact, the ratio of shell and bulk viscoelasticity $\kappa$ and interfacial shell viscosity $\eta_{\text{shell}} = \kappa \eta_{\text{macro}}$ scale for all tracer particle radii with the bulk viscosity $\eta_{macro}$ by the same power law as the averaged values shown in Figure 4D,E in the main text, which justifies the averaging approach in the main text.

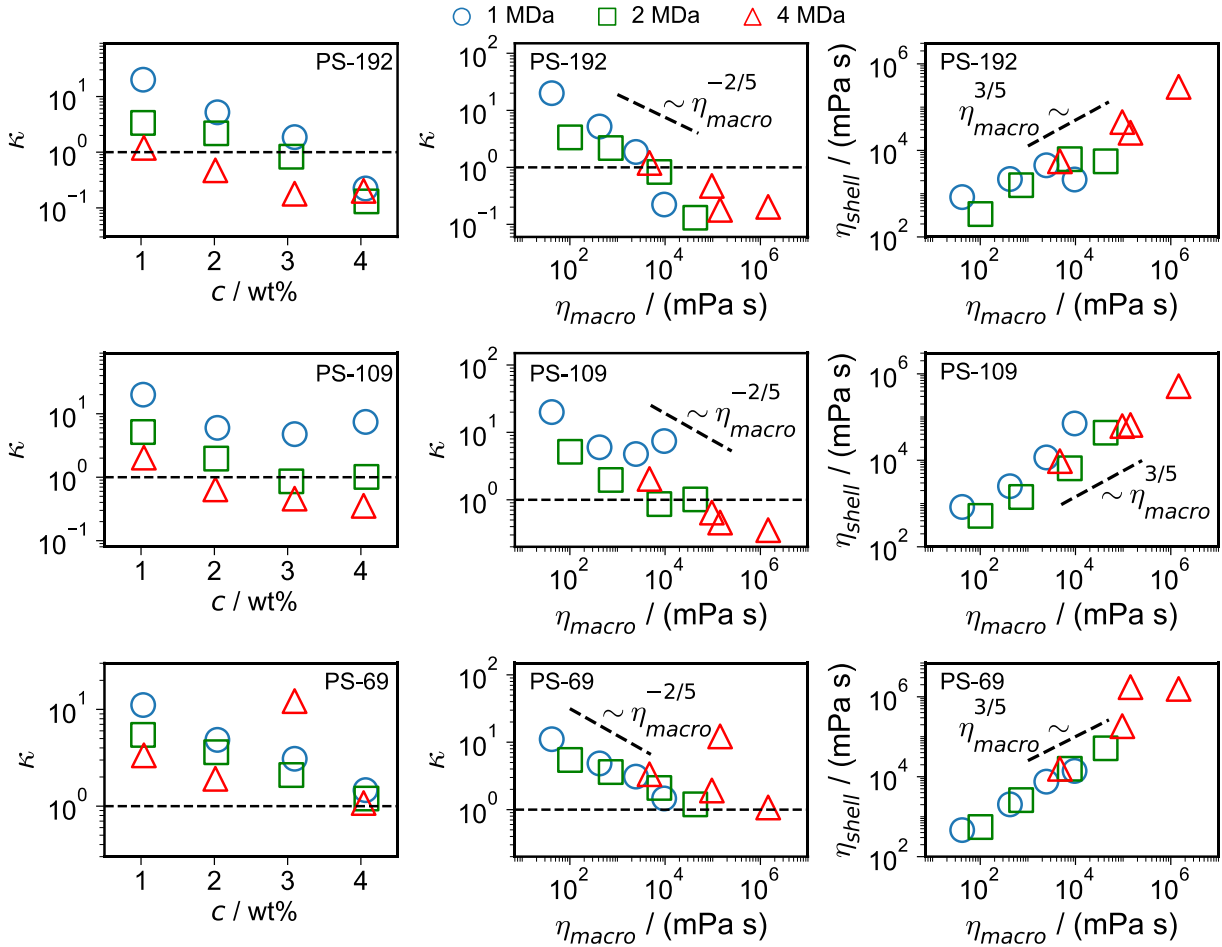

**Figure S23.** Ratio of shell and bulk viscoelasticity $\kappa = G^*_{shell}(\omega)/G^*_{macro}(\omega)$, which follows from the shift factor $\gamma_s$ in Figure 3C in the main text, as a function of polymer concentration and as a function of the bulk viscosity $\eta_{macro}$, for different tracer-particle radii and PEO molecular weights. Additionally, we show the interfacial shell viscosity $\eta_{shell} = \kappa \eta_{macro}$ in dependence of bulk viscosity $\eta_{macro}$. Power laws are added as guides to the eye.

## S19: Specification of particles

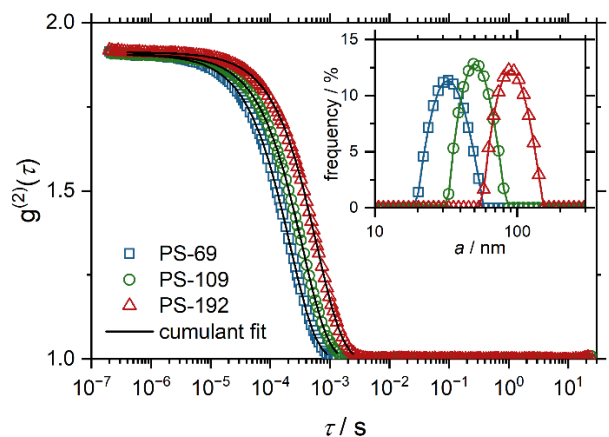

**Figure S24**. The radii of the polystyrene tracer particles were determined by dynamic light scattering and cumulant analysis.

Dynamic light scattering (DLS) measurements were performed on dilute aqueous solutions of all tracer particles. The tracer radius $a$ was determined by cumulant analysis of the intensity auto-correlation function provided by the Anton Paar Litesizer 500 instrument. The DLS data are shown in Figure S24, and the determined radii are shown in Table S3.

**Table S3.** Hydrodynamic radii $a$ and polydispersity index (PDI) of the tracer particles as determined by DLS.

| Abbreviation | Material | Surface Modification | $a$ / nm | Polydispersity Index / % |
|---|---|---|---|---|
| PS-69 | Polystyrene | None | 34.41 | 3.93 |
| PS-109 | Polystyrene | None | 54.65 | 1.86 |
| PS-192 | Polystyrene | None | 96.01 | 3.10 |

## S20: Amplitude sweep results

For all twelve samples, amplitude sweeps were performed at 6.3 rad/s. The results are shown in Figure S25. At a frequency of 6.3 rad/s = 1 Hz and an amplitude of 5%, the maximum shear rate is $\dot{\gamma}$ = 0.63 s$^{-1}$.

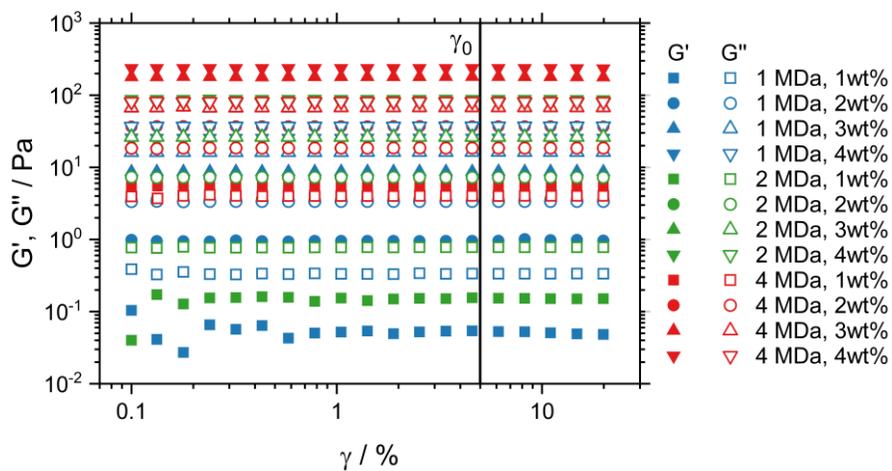

**Figure S25**. Amplitude sweep results for all twelve PEO samples. In all cases, the linear viscoelastic regime (LVE) extends to strain amplitudes higher than the maximum tested 20%. The chosen strain amplitude of $\gamma_0$ = 5% thus lies within the LVE for all samples.